\documentclass[aps,prb,twocolumn,amsmath]{revtex4}

\usepackage{graphicx}

\def\der#1#2{{\partial#1 \over \partial#2}}
\def\be{\begin{equation}}
\def\ee{\end{equation}}

\def\H{{\mathcal{H}}}

\def\Sz#1{\hat{S}^{z}_{#1}}
\def\Sx#1{\hat{S}^{x}_{#1}}
\def\Sy#1{\hat{S}^{y}_{#1}}
\def\O{\Omega}
\def\G{\Gamma}
\def\ba#1{\begin{array}{#1}}
\def\ea{\end{array}}
\def\r{\right}
\def\l{\left}
\def\summ{\sum\limits}
\def\prodd{\prod\limits}
\def\intt{\int\limits}
\def\b{{\bf b}}
\def\bd{{\bf b}^{\dagger}}

\begin{document}
\def\c{{\bf c}}
\def\cd{{\bf c}^{\dagger}}
\def\ket#1{\l|#1\r\rangle}
\def\lb{\label}
\def\nn{n}
\def\f{g}
\def\den{f}

\title{Transverse Meissner Physics of Planar Superconductors with
  Columnar Pins}
\author{Gil Refael$^1$, Walter Hofstetter$^2$, David R. Nelson$^3$}
\affiliation{$^1$ Department of Physics, California Institute of
  Technology, MC 114-36, Pasadena, CA 91125\\
$^2$ Institut f\"ur Theoretische Physik, Johann Wolfgang Goethe-Universit\"at, 60438 Frankfurt am Main, Germany\\
$^3$ Department of Physics, Harvard University, Cambridge MA, 02138}


\date{\today}

\begin{abstract}

The statistical mechanics of thermally excited vortex lines with
columnar defects can be mapped onto the physics of interacting quantum
particles with quenched random disorder in one less dimension. The
destruction of the Bose glass phase in Type II superconductors, when
the external magnetic field is tilted sufficiently far from the column
direction, is described by a poorly understood non-Hermitian quantum
phase transition. We present here exact results for this transition
in (1+1)-dimensions, obtained by mapping the problem in the hard core
limit onto one-dimensional fermions described by a non-Hermitian tight
binding model.   Both site randomness and the relatively unexplored
case of bond randomness are considered.  Analysis near the mobility
edge and near the band center in the latter case is facilitated by a
real space renormalization group procedure used previously for
Hermitian quantum problems with quenched randomness in one dimension.

\end{abstract}

\maketitle

\section{Introduction \label{intro}}

The physical properties of vortices in random pinning potentials have
been the focus of investigation for many years. Recent progress
has taken place in a particular aspect of this problem - vortices
trapped inside a slab of superconducting material with columnar
defects. In general, the problem of vortex pinning in a superconductor
with columnar defects can be recast via the transfer matrix method in terms
of the non-Hermitian quantum mechanics of bosons  with a constant
imaginary vector potential.
\cite{NelsonVinokur1993,HatanoNelson1997} In the case of a thin
superconducting slab, the mapping to non-Hermitian quantum mechanics
leads to a one-dimensional problem. In this case there has been recent
progress in the following specific problems: (1) Vortex hopping in a
regular array of columnar pins with on-site and nearest-neighbor
repulsion. \cite{Hofstetter2003,Affleck2004} (2) Transmission through a
weak-link. \cite{Hofstetter2003,Affleck2004} (3) Vortex tunneling with
Cauchy-distributed random on-site pinning potential (Lloyd
model)\cite{Lloyd, Brouwer, Halperin} (4) Some progress in
understanding vortex dynamics with on-site and hopping energies both
random was made in
Ref. \onlinecite{ShnerbNelson1998,Dahmen,GoldsheidKhoruzhenko}.

In this paper we first concentrate on a fifth case, namely, random hopping
in an otherwise uniform pinning array. The approach we take here is
the strong randomness real-space renormalization group (RSRG). 
This approach, pioneered by Ma, Dasgupta, and Hu,\cite{MaDas1979,MaDas1980} and developed further by Fisher, \cite{DSF94,DSF95} is very successful in treating
one dimensional random spin chains as well as other models.
\cite{Damle,Altmanetal2004,randomreview} Use of
this technique in the context of non-Hermitian transfer matrices,
however, appears to be new. We note that recently a related problem
of a vortex lattice pinned by a single columnar defect was considered
by Radzihovsky in Ref. \onlinecite{Radz}.

This problem is of interest for several reasons.  First,
although much is known about non-Hermitian tight binding models
with site randomness in (1+1)-dimensions,\cite{HatanoNelson1997,Brouwer,Halperin, ShnerbNelson1998,Dahmen,GoldsheidKhoruzhenko} the
case of pure bond randomness (corresponding to irregularly
spaced columnar pins of equal strength in a slab) is
relatively unexplored. All states are localized for the
Hermitian problem of vortices in a thin slab without an
external tilt field when only site randomness is present.
The tilt field $h$ must then exceed a finite threshold before
delocalized states appear in the center of the band.\cite{HatanoNelson1997}  In
contrast, there is always one delocalized state exactly at the
band center for the Hermitian problem with only random
hopping, accompanied by a diverging localization length of the
localized states on either side.\cite{Eggarter}

\begin{figure}
\includegraphics[width=8.5cm]{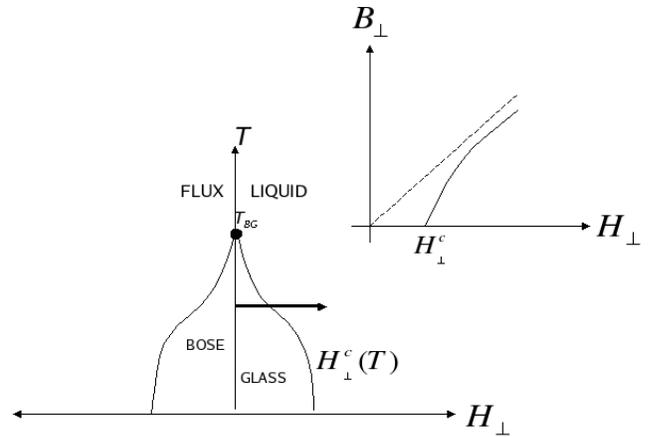}
\caption{Schematic phase diagram of a flux liquid with columnar pins
  at a fixed external magnetic field $\vec{H}_{||}$ parallel to the pins as a function of temperature T
  and tilted external field $H_{\perp}$.  The transitions along the
  line  $H_{\perp}^c(T)$  map onto the
  non-Hermitian quantum phase transition problem discussed in this
  paper. Inset shows the behavior of $B_{\perp}$ near $H_{\perp}^c$, where the transverse
  Meissner effect breaks down and the vortices begin to tilt.
\label{figA}}
\end{figure}

As we show explicitly in Sec. \ref{numerics}, additional delocalized states then
appear immediately in the center of the band for any nonzero tilt in
the thermodynamic limit.    As discussed further below, for larger
values of $h$ both problems do have similar mobility edges, separating
localized states near the band edges from delocalized states near the
band center.

\begin{figure*}
\includegraphics[width=12cm]{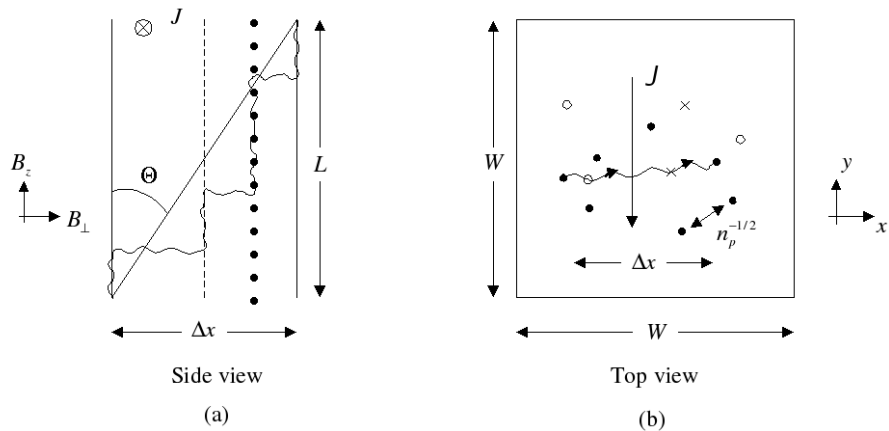}
\caption{(a) Schematic of a single tilted vortex line in a sample of
  thickness L and inclined at an average angle $\theta$, interacting
  with a row of variable strength columnar pins (represented by
  solid dashed and dotted vertical lines) used in our
  estimate of the flux flow resistivity in the presence of a current J
  flowing into the plane of the diagram.  (b)  Top view of this same
  situation, showing the path of the tilted vortex projected down the
  z-axis. $n_p$ is the density of columnar pins. 
  \label{figB}}
\end{figure*}

Perhaps more important, the analytically tractable free fermion model
discussed here sheds light on the poorly understood non-Hermitian
quantum phase transition that describes the physics of vortices as one
attempts to tilt them away from the direction preferred by columnar
defects.   Suppose for simplicity the longitudinal applied
magnetic field (i.e., $H_{||}$, the field parallel to the columnar pins)
produces a vortex density which does not exceed the density of column
pinning sites.   Then, as shown in Fig. \ref{figA}, the low temperature Bose
glass phase, with essentially all flux lines localized on columnar
defects, is expected to be stable for a small additional field $H_{\perp}$
perpendicular to the column direction.  The transverse applied field $H_{\perp}$
is proportional to the tilt field $h$  discussed in this paper.  Below the
zero tilt Bose glass transition temperature $T_{BG}$, perpendicular fields
less than a critical value $H^c_{\perp}$ leave vortices untilted and trapped on
columnar pins in the thermodynamic limit.\cite{NelsonVinokur1993}  The equivalent quantum
problem involves interacting quantum bosons in a disorder potential
and constant imaginary vector potential proportional to $H_{\perp}$.\cite{HatanoNelson1997}
Provided the transition is not first order, the breakdown of this
transverse Meissner effect above $H_{\perp}^c$ can be described by a critical
exponent $\zeta$, according to 
\be
B_{\perp}\propto\l(H_{\perp}-H_{\perp}^c\r)^{\zeta},
\lb{a}
\ee
where $B_\perp$ is the transverse flux due to the tilted vortices. 
Heuristic random walk arguments based on
the entropy of vortices wandering in the presence of thermal
fluctuations lead to the estimates,\cite{HwaNelsonVinokur} 
\be
\ba{c}
\zeta=\frac{3}{2},\,(d=3)\vspace{2mm}\\
\zeta=\frac{1}{2},\,(d=2)
\ea
\lb{b}
\ee
in three and two dimensions respectively.  However, there are reasons
to doubt these predictions.   First, as will be become clear, at least
for the exactly soluble (1+1)-dimensional model discussed in this
paper, the non-Hermitian quantum phase transition induced by tilt
effectively occurs at a finite wave-vector, calling into question
simple random walk arguments based on physics at $k=0$.  Second,
although the phase diagram shown in Fig. \ref{figA} is well established
experimentally,\cite{Yeh,Grigera,Olive}  the one existing experimental measure of the
exponent $\zeta$ \cite{Crabtree} (based on current-voltage characteristics, see below) is
inconsistent with Eq. (\ref{b}). Indeed, the experiments on bulk
superconductors with columnar pins in Ref. \onlinecite{Crabtree} find $\zeta=1/2$, which disagrees
with the prediction of Eq. (\ref{b}) in $d=3$.  In fact, we show in this paper
that, for the special case of  free fermions in (1+1)-dimensions
(corresponding to Luttinger liquid parameter $g=1$ in Fig. \ref{phasediagram}) one obtains 
\be
\zeta=1 \ \ \ \mbox{(free fermions, 1+1 dimensions)},
\lb{c}
\ee
which disagrees with the prediction of Eq. (\ref{b}) in $d=2$.   

In principle it might be possible to check predictions like
those in Eqs. (\ref{b}) and (\ref{c}) by magnetic torque measurements,
which are sensitive to the differences in the direction of
$\vec{B}$ and $\vec{H}$ that are an essential part of the transverse Meissner
effect.\cite{Grigera}  Alternatively, as discussed in
Refs. \onlinecite{NelsonVinokur1993} and \onlinecite{HwaNelsonVinokur},
the exponent $\zeta$ determines a density of kink excitations
connecting nearby columnar defects (see Fig. \ref{figB}), which in
turn control the linear flux flow resistivity above $H_{\perp}^c$. 
Current-voltage curves ar highly nonlinear below $H_{\perp}^c$,  and
the linear resistivity vanishes.\cite{NelsonVinokur1993} To
understand the linear resistivity above $H_{\perp}^c$, consider the
geometry shown in Fig. \ref{figB}a, where a current flows
perpendicular to the plane defined by the column direction and
the average field direction defined by a set of tilted vortex
lines. Imagine first a bulk sample, of dimensions $L\times W\times W$, where $L$
is the sample length along the columns. Suppose this field
is inclined at a small angle $\theta=B_{\perp}/B_{||}$ away from the column
direction. Then a typical perpendicular distance $\Delta x$ traversed by
a vortex  across a sample is $\Delta x=\theta L$.    If the density of columnar pins is $n_p$, this tilt leads to 
\be
N_1=\Delta x/n_p^{-1/2}=\frac{B_{\perp}}{B_{||}}n_p^{1/2} L
\lb{d}
\ee
kinks associated with a single vortex. In clean Type II
superconductors, at temperatures high enough so that residual pinning
by point impurities can be neglected, these kinks will slide along
columns, due to the Lorentz force $f_L=\phi_0 J/c$ caused by the
current, where $\phi_0$ is the
flux quantum and $c$ the speed of light. Upon multiplying by the total
number of flux lines $W^2 B_{||}/\phi_0$, where, we find a gas of kinks with density
\be
n_k=B_{\perp}n_p^{1/2}/\phi_0
\lb{e}
\ee
per unit volume.  As expected, $n_k$ is proportional to $B_{\perp}$,
which leads using Eq. (\ref{a}) to a flux flow resistivity
\be
\rho=\rho_0 B_{\perp}\xi^2/\phi_0\propto \l(H_{\perp}-H_{\perp}^c\r)^{\zeta},
\lb{f}
\ee
where $\rho_0$ is the normal state resistivity, $\xi$ (the coherence length) is the
size of the normal vortex cores which contribute to the dissipation,
and we have neglected coefficients of order unity.    A similar
calculation can be carried out for a two dimensional slab with
dimensions $L\times W\times d$, with $d\ll L,\,W$  and a single sheet
of columnar pins along $L$ with spacing $n_p^{-1}$.
We also require $d<\lambda$, 
where $\lambda$ is the London penetration depth.   We then find
that the flux flow resistivity for a current perpendicular to the slab
(i.e., along the direction $d$) is identical in form to the three
dimensional result (\ref{f}).

We hope that the calculations in this paper, although only valid in
(1+1) dimensions at a special temperature deep within the Bose glass
phase (see Fig. \ref{phasediagram}), will stimulate further theory as well as
experiments aimed at a more complete determination the exponent
$\zeta$, in both 2 and 3 dimensions.

\subsection{Statement of the problem}

When a magnetic field is applied to a planar superconductor with
parallel columnar defects in the plane, a competition occurs between the tendency of
the vortices to be pinned by the columnar defects, and the portion of
the in-plane magnetic field which is normal to the columnar defects. 
The resulting lock-in of vortex trajectories parallel to the columns
up to a critical tilt field is the transverse Meissner effect
discussed above.

In this work we concentrate on a model of vortex hopping in
a random array of columnar defects (see Fig.~\ref{columnar}). 
As described previously \cite{NelsonVinokur1993,HatanoNelson1997}, 
this statistical mechanics problem is mapped onto a
quantum mechanical boson-hopping problem. When the magnetic
field that produces the vortices is tilted relative to the columns, our model
introduces non-Hermiticity to the Hamiltonian. The Hamiltonian whose
exponential gives the transfer matrix is:
 \be
\ba{c}
\H=\summ_i \l(-w_i\l(\bd_i \b_{i+1} e^{-h_i}+\bd_{i+1} \b_{i}
e^{h_i}\r)\r.\vspace{2mm}\\\l. + \l(\epsilon_i - \mu \r) \bd_i\b_i
+ {U \over 2} n_i\l(n_i-1\r)\r).
\label{1.1}
\ea
\ee
Each vortex is represented by a boson with creation and annihilation
operators $\b, \bd$. As discussed above, we restrict our attention to {\it  hard core} vortices, 
and set the on-site repulsion to infinity ($U\rightarrow\infty$).

The average number $n$ of vortices per pinning site is equivalent to the \emph{longitudinal magnetization} 
induced by a magnetic field $H_{||}$ parallel to the defects. 
The part of the magnetic field $H_{\perp}$, tilted relative to the columnar defects, induces an  
imaginary vector-potential $h_i$ on each bond.  
We expect it to depend linearly on the distance between columnar pins. 
Note, however, that using a similarity transformation one
can redistribute the $h_i$'s such that each bond carries the same  
imaginary vector-potential \cite{ShnerbNelson1998,Dahmen}. 
In the following we will therefore eventually apply a uniform tilt $h$ for all bonds. 
In Eq. (\ref{1.1}), $\mu$ is the vortex chemical potential, which is controlled
by the external magnetic field and the depth of the pinning
potential. The hopping energies $w_i$ and pinning energies
$\epsilon_i$ are random variables with some
relatively well behaved distribution. As is shown schematically in
Fig. \ref{columnar}, the Hamiltonian (\ref{1.1})
describes the statistical mechanics of vortices hopping between 
columnar defects with an external magnetic induction tilted relative
to the columnar pins.
\begin{figure}
\includegraphics[width=8cm]{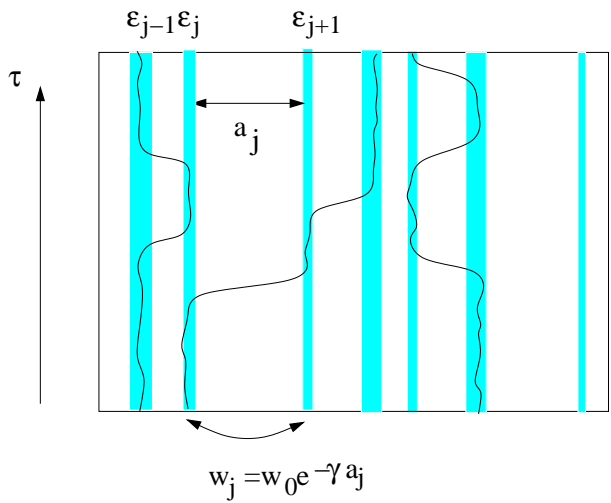}
\caption{The exponential of the Hamiltonian (\ref{1.1}) determines the
  transfer matrix describing the statistical
  mechanics of vortices (black lines) fluctuating in a planar superconductor
  with parallel columnar defects piercing it at random intervals (gray
  stripes). The distance between the defects, $a_j$, as well as their energy
  depth (illustrated by their width), $V_j$, are random
  variables. The irregular spacing leads to strong randomness in the vortex hopping,
  $w_j\propto e^{-\gamma a_j}$. The external magnetic induction and
  its direction relative to the columnar pins determine the vortex
  chemical potential, $\mu$, and the effective tilt $h_j\propto a_j$. 
In this paper we compare and contrast the case of identical pins with
  random spacings, and the case of random pinning energies but uniform spacings. 
  }
\lb{columnar}
\end{figure}

In general, our goal is to characterize the transverse Meissner effect
in the superconductor. For that purpose we also need to define the
transverse magnetic flux in terms of the bosons of
Eq. (\ref{1.1}). Quite intuitively, the transverse flux is equivalent to the
boson {\it current}, and is found by differentiating the Hamiltonian with
respect to the external transverse field: \cite{NelsonVinokur1993,HatanoNelson1997}
\be
{\bf J}_i=(-i) \der{\H}{h_i}= (-i) w_i\l(\bd_i \b_{i+1} e^{-h_i}-\bd_{i+1} \b_{i}
e^{h_i}\r).
\label{current_op0}
\ee

We will consider and contrast two cases: first, the case of random
  pinning energies with
  uniform spacings, and then the case of identical pins but random spacings. 
We will show that these two cases are quite different, and proceed to
  characterize the case of random spacings in some detail.

As we show in Sec. \ref{fermi}, the Hamiltonian (\ref{1.1}) with the
hard-core requirement ($U\rightarrow\infty$) is equivalent to a
fermion model without interactions. In general these interactions are quantified by the Luttinger parameter\cite{Affleck2004} 
\be
g\sim \frac{\pi T}{\sqrt{C_{11} C_{44}}},
\lb{g}
\ee
where $T$ is the temperature, $C_{11}$ and $C_{44}$ are the vortex
compressibility and tilt modulus respectively. In this work we
concentrate on the line $g=1$, and ignore nearest neighbor and higher
range interactions between the
hard-core vortices. The general phase diagram for the interacting
vortex lines with tilt in $(1+1)$-dimensions is
shown in Fig.~\ref{phasediagram}. \cite{NelsonVinokur1993,HwaNelsonVinokur}

\begin{figure}
\includegraphics[width=8cm]{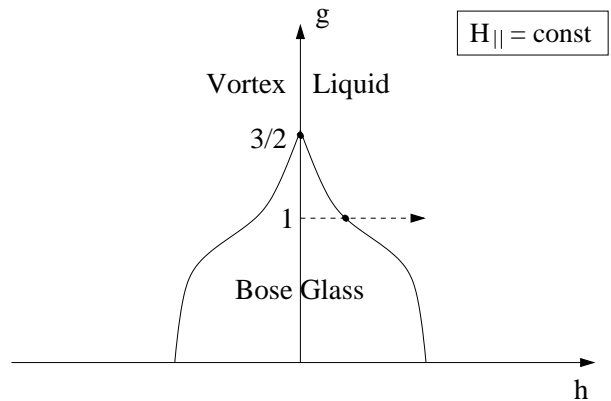}
\caption{
The phase diagram for interacting vortices, with effective
  Luttinger parameter $g$ proportional to temperature (see Eq. \ref{g}), tilt $h$, at a fixed parallel
  magnetic field $H_\parallel$. For real bosons, the vortex liquid phase for
  $g>3/2$ and vanishing tilt $h=0$ corresponds to a superfluid with off-diagonal long
  range order. Little is known about the
  Bose-glass depinning transition at finite $h$. In this paper we investigate this transition at $g=1$.}
\lb{phasediagram}
\end{figure}

\subsection{Summary of results}

In the first part of this paper (Secs. \ref{sec2} and \ref{sec3}) we review and derive properties of the
random-pinning Lloyd model. We concentrate on the minimum tilt required to form
delocalized eigenstates, i.e., the critical tilt as a function of parallel
field to destroy the transverse Meissner effect, and the
resulting transverse flux as a function of both tilt and
parallel field. The derivation of these properties provides us with a
baseline for a comparison of the random pinning model with the
random-hopping model. 

In the second part of the paper we concentrate on random-hopping. In
this model all pinning sites are assumed to be identical,
but with random distances between them (since hopping depends
exponentially on the inter-site distance, the hopping strengths, $w_i$,
will be strongly random). For this purpose we employ the real-space renormalization group 
(RSRG) method. Using this method we obtain the density of states for the vortex-hopping
Hamiltonian. We derive this method for the vortex-hopping problem in
Sec. \ref{RSRGsec}. 

The random-hopping problem has two unique features that are connected
to each other: the localization length of the vortex eigenfunctions
diverges near the middle of the band ($E=0$), and there is also a
singularity of the DOS at the same place.  We use the DOS to derive a
relationship between  the effective vortex chemical potential, $\mu$,
and the applied parallel field $B_{\parallel}$. We then proceed to show that any
nonzero tilt will produce delocalized eigenstates, and derive the
critical tilt $h_c(b)$ at which the transverse Meissner effect breaks down. We
also derive the localization length of vortex states near the mobility
edge.

By employing a simple spectral formula explored in
Refs. \onlinecite{Brouwer}  and \onlinecite{ShnerbNelson1998,Dahmen}, together with the
results of the real-space RG and the general arguments in 
App. \ref{appcurrent}, we derive the following properties: the angle of
approach of the delocalized spectrum in the complex plane ($d{\rm Im}
E/d{\rm Re} E$), the contribution to the transverse
magnetization of a single delocalized vortex state, and the total
vortex current, or transverse flux, near the breakdown of the
transverse Meissner effect. 

We emphasize that the results derived here assume that the system is
in the universal low energy limit. A numerical investigation of this limit requires large system
sizes that allow the RSRG to reach low energies.  This is confirmed in Sec.~\ref{numerics} 
where finite-size systems of vortices and columnar pins with 
random pinning and random hopping are diagonalized exactly.

\section{Equivalent models \label{sec2}}

Hamiltonian (\ref{1.1}) in the limit $U\to\infty$ is not as familiar to
us as some other equivalent models. In this section we map 
the boson Hamiltonian (\ref{1.1}) to a spin model with easy plane
anisotropic interactions (XX), and to a fermion hopping model. 
These mappings will be useful when applying the RSRG in Sec.~\ref{RSRGsec}   
and exact diagonalization for finite systems in Sec.~\ref{numerics}.

\subsection{Mapping to a spin model}

By the simple transformation of the boson operators on the n$^{th}$ 
lattice site,
\be
\ba{c}
\bd_n=(-1)^n\cdot\hat{S}^+_n=(-1)^n\cdot\l(\Sx{n}+i\Sy{n}\r),  \vspace{2mm}\\
  \b_n=(-1)^n\cdot \hat{S}^-_n=(-1)^n\cdot\l(\Sx{n}-i\Sy{n}\r), \vspace{2mm}\\  
  \bd_n\b_n=1/2+\Sz{n},
\ea
\label{1.2}
\ee
the Hamiltonian in Eq. (\ref{1.1}) is transformed to an XX
ferromagnet in an external magnetic field:
\be
\ba{c}
\H= \summ_i \l(-2w_i\l(\l(\Sx{i}\Sx{i+1}+\Sy{i}\Sy{i+1}\r)\cosh(h)\r.\r.\vspace{2mm}\\\l.\l.
+i\l(\Sx{i}\Sy{i+1}-\Sy{i}\Sx{i+1}\r)\sinh(h)\r)
\r.\vspace{2mm}\\\l.
+ \l( \epsilon_i - \mu \r) \l(\frac{1}{2}+\Sz{i}\r)\r).
\label{1.3}
\ea
\ee
In the spin variables, a vortex pinned at site $i$ corresponds to
$\Sz{i}=+1/2$; an empty site $i$ is transformed to a site with
$\Sz{i}=-1/2$. The transformation in Eq. (\ref{1.3}) is the special
spin-1/2 case of the Holstein-Primakoff transformation.\cite{Holstein-Primakoff} In its higher
spin version it has factors of the form $\sqrt{1-\bd_n\b_n}$ which for
spin-1/2 reduce to either zero or one. 

\begin{figure}
\includegraphics[width=8cm]{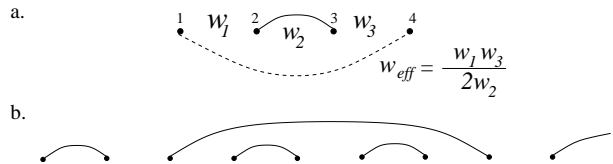}
\caption{(a) In a random XX spin chain, described by Eq.~(\ref{1.3}) with $h=0$, strong bonds
  such as $w_2$ localize a spin-singlet. Quantum fluctuations induce a
  coupling between the neighbors of the singlet. This coupling has the
  same XX form, but a substantially reduced scale $w_{\rm eff}=w_1w_3/w_2$ which
  is much smaller than $w_1,\,w_2$, and $w_3$. By repeating the
  singlet formation process, the energy scale of the effective
  Hamiltonian is reduced. (b) The random singlet state. Singlets form
  in a random fashion, mostly between nearest neighbors, but they also
  connect very far away spins. Long distance singlets give rise to
  average correlations that decay algebraically with
  distance. \label{fig1}}
\end{figure}

At half-filling (i.e., when $\mu=\epsilon_i=0$) and zero tilt $h$ the ground state of Hamiltonian (\ref{1.3}) is the random
singlet phase. In this phase singlets form in a random fashion
between sites connected by strong $w_i$. Once a singlet forms, the nearest neighbors of the two sites
involved also interact, but with suppressed strength. By pairing up
strongly-interacting sites into singlets, we  iteratively reduce the
energy scale of the Hamiltonian, until we exhaust all sites. At this
point, singlets connect many nearest neighbors, but occasionally
singlets connect very far away spins, leading to power-law decaying
correlations (for a review, see Ref. \onlinecite{DSF94}). The idea
behind the random singlet
phase and its formation are shown in Fig.~\ref{fig1}. A thorough
discussion of this phase and its implications for vortex pinning will
be given in Secs. \ref{RSRGsec} and \ref{conn}.

\subsection{Mapping to fermion hopping problem \lb{fermi}}

By using a version of the Wigner-Jordan transformation,
\cite{Wigner-Jordan} we can also map the
hard-core boson Hamiltonian, Eq.~(\ref{1.1}) with $U=\infty$, to a fermionic random
hopping problem. This mapping is also quite straightforward; we start
with
\be
\ba{cc}
\b_n\rightarrow \prodd_{j=-\infty}^{n-1} e^{i\pi \cd_j\c_j} \c_n,  &
\bd_n\rightarrow \prodd_{j=-\infty}^{n-1} e^{-i\pi \cd_j\c_j} \cd_n. 
\ea
\label{1.6}
\ee
where the string operator ensures the anti-commutation relations of the
$c_n$'s. Now we can write the Hamiltonian of the vortices as though
they are fermions:
\be
\H=\summ_i \l(-w_i\l(\cd_i \c_{i+1} e^{-h_i}+\cd_{i+1} \c_{i}
e^{h_i}\r) + \l( \epsilon_i - \mu \r) \cd_i\c_i\r).
\label{1.7}
\ee
Similarly, the local current operator becomes:
\be
\ba{c}
{\bf J}_i=(-i) \der{\H}{h_i}\vspace{2mm}\\
=(-i) w_i\l(\bd_i \b_{i+1} e^{-h_i}-\bd_{i+1} \b_{i}
e^{h_i}\r)\vspace{2mm}\\
=(-i) w_i\l(\cd_i \c_{i+1} e^{-h_i}-\cd_{i+1} \c_{i}
e^{h_i}\r).
\label{current_op}
\ea
\ee

In the absence of the tilt field, the ground state of the Hamiltonian
(\ref{1.7}) is well understood.\cite{DSF94} When the pins are identical ($\epsilon_i = 0$), 
the random hopping localizes all states {\it except} at half filling, where there is always a
delocalized state. The localized states
away from the mobility edge are related to the random singlets in the
XX spin chain of Eq. (\ref{1.3}). 
Instead of a singlet, however, pairs of sites share a single fermion. 

In this model, half filling is obtained when $\mu \to 0^-$. In this
limit, the last fermion inserted in the system is delocalized
between two sites with a distance that is of the order of the system size, i.e., it
is delocalized. All other fermionic states, however, are localized.

In the presence of the tilt field, $h$, an entire band of delocalized
states appears, and the lower mobility edge moves down
to fillings below one half, and to negative chemical potentials $\mu<0$. By using
the mapping to free-fermions (Eq. \ref{1.7}) and the known real space
renormalization group (RSRG)
results for this model, we can obtain much insight
into the delocalized phase. In particular, we will derive a universal 
relationship between the field $h$, the mobility edge $\mu_h$, and
perhaps most importantly, the
vortex density (i.e. parallel magnetic field) at the mobility edge, $\rho_h$. 


\section{Random pinning energy and uniform hopping - exact results
  from the Lloyd model \label{sec3}}

In this section we will  review exact results for the Lloyd model. \cite{Hofstetter2003,Lloyd, Brouwer}
We consider the one-dimensional quantum Hamiltonian, Eq. (\ref{1.1}) 
with infinite on-site repulsion ($U\to \infty$):
 \be
\H=\summ_i \l(-w\l(\bd_i \b_{i+1} e^{-h}+\bd_{i+1} \b_{i}
e^{h}\r) - \l(\epsilon_i - \mu \r) \bd_i\b_i\r).
\lb{w.1}
\ee
where $\bd_i$ and $\b_i$ are creation and annihilation operators of
hard-core bosons, which represent the vortices. The hopping matrix
element $w$ is now site independent. We could equally well use the
fermion representation in Eq. (\ref{1.7}) in what follows. We set the lattice
constant to $1$, and consider a lattice of length $L$ sites. 
For small tilts, the imaginary gauge-field $h$ is proportional to the angle of the
applied magnetic field relative to the columnar defects. The parallel
applied field translates into the chemical potential for the vortices
via the constraint,
\be
\intt_{-\infty}^{\mu} \f(\epsilon)d\epsilon=\nn,
\lb{w.1.5}
\ee
where $\nn$ is the average number of vortices per columnar defect, and
$\f(\epsilon)$ is the density of states associated with the
Hamiltonian of Eq. (\ref{w.1}). Assuming that the distribution of
pinning energies, $\epsilon_i$, is symmetric, we can
invoke particle hole symmetry, and rewrite Eq. (\ref{w.1.5}) as:
\be
\intt_{\mu}^0 \f(\epsilon)d\epsilon=\intt_{-\infty}^0 \f(\epsilon)d\epsilon-\intt_{-\infty}^{\mu} \f(\epsilon)d\epsilon=0.5-\nn.
\lb{w.1.51}
\ee

In this section we will review the known qualitative feature of this
non-interacting model of vortices with random pinning strength $\epsilon_i$. 
We will also focus on the analytic
results known for the Lloyd model,\cite{Halperin,Lloyd, Brouwer} which is described by the
Hamiltonian (\ref{w.1}) with a special distribution of the pinning
energies  $\epsilon_i$:
\be
P[\epsilon]=\frac{\gamma}{\pi}\frac{1}{\epsilon^2+\gamma^2}
\lb{w.2}
\ee
This model
has played an important role in our understanding of one-dimensional
localization of electrons.

\subsection{Structure of the spectrum and critical delocalization tilt}

Much of the physics of interacting vortices in $(1+1)$-dimensions can be inferred from the
shape of the spectrum of the Hamiltonian as a function of the applied transverse field $h$. 
For zero tilt $h$, the entire spectrum of Hamiltonian
(\ref{w.1}) is due to localized states, and is therefore real.\cite{HatanoNelson1997}
The density of states changes as a function of energy but experiences no singularities. When 
the external magnetic field is tilted relative to the columnar
defects, the spectrum of the localized states does not change, but the
wave functions associated with them spread out in the direction of the
tilt. The invariance of the spectrum to the tilt, as long as the
wave-functions are localized, can be easily understood, since for any
eigenstate of the Hamiltonian we can carry out a  gauge transformation
such that
all the imaginary gauge-field in a sufficiently large system is on a bond between sites where
the eigenstate has no support. 

\begin{figure}
\includegraphics[width=8cm]{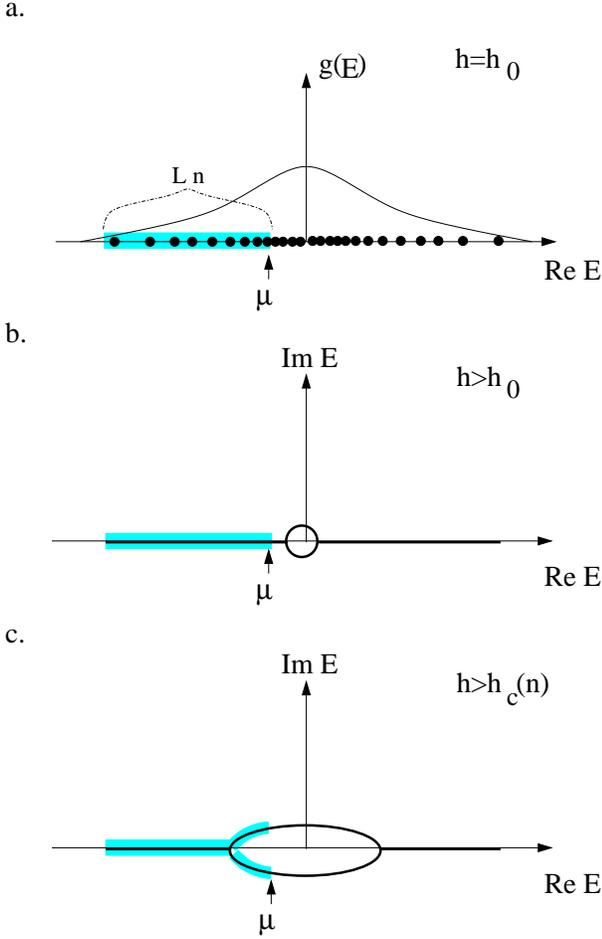}
\caption{(a) All eigenfunctions of the random-pinning Hamiltonian are
  localized and independent of the tilt provided  $h<h_0$. 
  We suppose that the density of states (DOS) is peaked around
  $E=0$ but is not singular. The effective chemical potential of the
  vortices, $\mu$, is determined by the longitudinal magnetization $n$, 
  such that the density of occupied states below $\mu$
  equals $n$. (b) When the tilt of the applied field becomes $h>h_0$,
  a bubble of delocalized states develops where the DOS is peaked. As
  long as the bubble does not reach $\mu$, there is still no transverse
  magnetization. The thick lines mark the support of the
  eigen-energies in the complex E-plane. (c) The transverse Meissner effect breaks down when
  $h=h_c$, and the delocalized bubble reaches $\mu$. The {\it total} vortex
  current is shown in App.~\ref{appcurrent} to be proportional to the
  imaginary part of the energy eigenvalue at the Fermi surface
  (c.f. Eq. \ref{7.8}). 
\label{spectrum}
}
\end{figure}

Above some critical tilt $h_0>0$, a subset of the eigenfunctions 
delocalize, and their energies become complex (see Fig. \ref{spectrum}). As shown in
Ref. \onlinecite{ShnerbNelson1998,Dahmen}, the parts of the spectrum that
first become delocalized are at the maximum of the DOS
$\f(\epsilon)$, which is generally in the center of the spectrum, at $\epsilon=0$.  

The transverse Meissner effect, however, persists until the
vortex-eigenstates at the ``Fermi energy'' $\mu$ become delocalized. 
This Fermi energy for the vortices is determined by the
parallel (longitudinal) magnetic field $B_{\parallel}$, or equivalently by the vortex density $n$ per pin.  
The tilt at which the transverse Meissner effect breaks
down is thus a function of $\nn$, and it is always bigger than the
threshold-tilt $h_0$: 
\be
h_c(\nn) \ge h_0. 
\lb{w.3}
\ee

The qualitative description above is exemplified by the 
spectrum of
the the Lloyd model.\cite{Lloyd, Brouwer} 
The delocalized portion of the spectrum of the Lloyd model is known to be:\cite{Lloyd}
\be
E_n^{\pm}=-2w\cos(\pm k_n+ih)\mp i\gamma,
\lb{8.3}
\ee
where $\gamma$ is controls the width of the site-randomness
distribution in Eq. (\ref{w.2}). As the tilt increases such that $h>h_0$, a bubble of delocalized states appears in the center of
the band (c.f. Fig. \ref{spectrum}). When $h>h_c(n)$, this bubble
engulfs the chemical potential.

From the spectrum of the Lloyd model, Eq. (\ref{8.3}), one can easily
derive $h_0$ and $h_c$. The lowest critical tilt $h_0$ that
produces delocalized eigenfunctions is obtained when the imaginary
part of the energy becomes nonzero, ${\rm Im}
E_n\neq 0$, for some eigenstate $E_n$. 
As is obvious from Fig. \ref{spectrum}b, this delocalization tilt will
be the critical tilt where the transverse Meissner effect disappears at
half filling ($n=0.5$). $h_0$ is given by:
\be
\sinh (h_0)=\frac{\gamma}{2w}
\lb{8.4}
\ee
The range of wave-vectors $k_n$ that defines the delocalized states are thus given by:
\be
\sin|k_n|>\frac{\gamma}{2 w \sinh(h)}
\ee
To obtain the critical tilt as a function of longitudinal magnetic
field, we observe that the number of delocalized single-vortex states is 
\be
N_{del}=\frac{2L}{\pi}\arccos\l( \frac{\gamma}{2w\sinh(h)}\r)
\lb{8.5}
\ee
(note that the argument of the $\arccos$ is never bigger than $1$). 
Thus the critical longitudinal vortex density for a given tilt is:
\be
\nn_c=\frac{1}{2}-\frac{1}{\pi}\arccos\l(\frac{\gamma}{2w\sinh(h)}\r),
\lb{8.6}
\ee
which can be inverted to give the critical tilt for a given magnetization:
\be
\sinh(h_c)=\frac{\gamma}{2w\cos(\pi(0.5-n))}.
\lb{8.7}
\ee
 
\subsection{Transverse magnetic flux near critical tilt}

Once vortices form delocalized states, the magnetization in the
superconductor is no longer parallel to the columnar defects, and
transverse flux appears.
The total transverse flux, or vortex current in the
quantum-mechanical picture,  can also be found out easily for the
Lloyd model.\cite{Lloyd}

In App. \ref{appcurrent} we derive a general rule for the total vortex
 current for delocalized states (Eq. \ref{7.8}):
\be
J_{total}=2\frac{1}{L}\summ_{k_n}{\rm Re}\der{E_n}{h}=\frac{1}{\pi}\l({\rm Im} E_{\mu}-{\rm Im} E_{\mu_m}\r)
\ee
where ${\rm Im} E_{\mu}$ is the imaginary part of the energy eigenvalue
with real part $\mu$ and $\mu_m(h)$ is the chemical potential above which 
vortex states are delocalized (mobility edge). $L$ is the total number
 of lattice sites. 
As discussed in Ref. \onlinecite{HatanoNelson1997,Affleck2004}, this
 total 'current' (similar to the derivative with respect to vector
 potential which gives the current in a quantum system) is
 proportional to the perpendicular component of the magnetic flux. If
 we consider our system having longitudinal vortex density $n$ and
 tilt $h>h_c$, then the number of {\it delocalized} vortices (as opposed to vortex states) is $\nn-\nn_c$ where $\nn_c$ is given in Eq. (\ref{8.6}).
Therefore, the $k_n$ that are occupied by delocalized vortices are
 given by 
\begin{widetext}
\be
\arcsin\l(\frac{\gamma}{2w \sinh(h)}\r)=k_c<|k_n|<\arcsin\l(\frac{\gamma}{2w \sinh(h)}\r)+\pi \l(\nn-\nn_c\r)=k_F
\lb{8.8}
\ee
Now, using Eq. (\ref{7.8}) (see Appendix A) we can directly write the total current as:

\be
\ba{c}
J_{total}=\frac{1}{\pi}\l({\rm Im} E_{k_F}-{\rm Im} E_{k_c}\r)=\frac{1}{\pi}2w\sinh(h)\cdot\l(\sin\l(\arcsin\frac{\gamma}{2w \sinh(h)}+\pi \l(\nn-\nn_c\r)\r)-\sin\l(\arcsin\frac{\gamma}{2w \sinh(h)}\r)\r)\vspace{2mm}\\
=\frac{1}{\pi}2w\sinh(h)\cdot\l( \frac{\gamma}{2w \sinh(h)}\l(\cos\l(\pi \l(\nn-\nn_c\r)\r)-1\r)+\sqrt{1-\frac{\gamma^2}{(2\sinh(h)w)^2}}\sin\l(\pi \l(\nn-\nn_c\r)\r)\r).
\lb{8.9}
\ea
\ee
\end{widetext}
Near the delocalization transition we have:
\be
J_{total}=2w\sinh(h)\sqrt{1-\l(\frac{\gamma}{2w\sinh(h)}\r)^2}(\nn-\nn_c).
\lb{8.10}
\ee
Using the derivative of Eq. (\ref{8.6}) we obtain 
\[
\frac{d\nn_c}{dh}=\frac{1}{2\pi}\frac{\gamma}{2w}\frac{\cosh(h)}{\sinh^2 h}\frac{1}{ \sqrt{1-\l(\frac{\gamma}{2w\sinh(h)}\r)^2}}
\]
Upon putting this back in Eq. (\ref{8.10}) with $h-h_c\approx
|d\nn/dh\cdot \l(\nn-\nn_c\r)|$  we obtain a simple expression for the
total current in the Lloyd model near delocalization, namely
\be
J_{total}\approx\frac{\gamma}{2\pi\tanh(h)}\cdot\l(h-h_c\r).
\lb{eq24}
\ee
This linear onset of the vortex current is qualitatively similar to numerical results 
for a box distribution of pinning energies, as presented in Section
\ref{numerics}. When reintegrated as a formula for the transverse
magnetic flux of a vortex system, Eq. (\ref{eq24}) leads to the
prediction $\zeta=1$ for the exponent discussed in the introduction,
Eq. (\ref{b}).


\section{Real space RG of the boson hopping problem \lb{RSRGsec}}

One purpose of this paper is to understand the properties of vortices
hopping in an array of identical pins with random locations, and contrast these
properties with those of the random-pinning energy system with uniform
hopping. The random-hopping
problem is described
by the Hamiltonian in Eq. (\ref{1.1}) with $U\to\infty$, $\epsilon_i=0$, and
$w_i$ random. In order to
solve this model we apply the real-space renormalization group (RSRG)
procedure.\cite{MaDas1979,MaDas1980,DSF94,DSF95} In this section we review the application of
the RSRG to the random hopping boson Hamiltonian.

The RSRG allows us to diagonalize the random Hamiltonian 
iteratively by eliminating the high energy degrees of freedom. 
The first step is finding the strongest bond in the chain and diagonalizing it while ignoring the rest of the Hamiltonian. 
Quantum fluctuations give rise to new interactions
that bridge over the bond we diagonalize. If we are lucky, these
renormalized interactions will be of the same form as the original
Hamiltonian. Let us carry this out for the boson-hopping Hamiltonian
Eq. (\ref{1.1}). In effect, we diagonalize the Hamiltonian
(\ref{1.1}) iteratively by putting vortices into a hierarchy of {\it localized}
states shared by pairs of sites.  

\subsection{Diagonalization of the strongest bond}

The Hamiltonian we are interested in is given in Eq. \ref{1.1}, with $U\rightarrow\infty$ and 
$\epsilon_i = 0$: 
\[
\H=\summ_i \l(-w_i\l(\bd_i \b_{i+1} e^{-h}+\bd_{i+1} \b_{i}
e^{h}\r)-\mu \bd_i\b_i\r).
\]
Suppose $w_i$ is the strongest hopping energy in the chain. We first
diagonalize a single nearest neighbor bond:
\be
\H_i=-w_i\l(\bd_i \b_{i+1} e^{-h}+\bd_{i+1} \b_{i}e^{h}\r)
- \mu (\bd_i\b_i  + \bd_{i+1} \b_{i+1})
\label{2.1}
\ee
Because the vortices are hard core in this limit the
only occupied states available in the Hilbert space are
\be
\ba{cc}
|i\rangle=\bd_i|0\rangle, \ \  \ |i+1\rangle=\bd_{i+1}|0\rangle.
\label{2.2}
\ea
\ee
Diagonalizing Eq. (\ref{2.1}) in the subspace spanned by
(\ref{2.2}) means diagonalizing the matrix:
\be
\l( \ba{cc}
-\mu & -w_i e^{-h} \vspace{2mm}\\
-w_i e^{h} & -\mu \ea \r)
\label{2.3}
\ee
Note that the eigenvalues of this Hamiltonian can not depend on the
tilt, in accordance with what we know on localized states - they carry
no transverse magnetic field.\cite{HatanoNelson1997} The eigenvalues and right eigenvectors of
(\ref{2.3}) are
\be
\ba{cc}
E_{\pm}=-\mu\pm w_i \vspace{2mm}\\
|\pm\rangle=\frac{1}{\sqrt{2}}\l(e^{-h/2}|i\rangle\mp e^{h/2}|i+1\rangle\r).
\ea
\label{2.4}
\ee
Note that in this non-Hermitian quantum problem, the left eigenvectors are not the Hermitian 
conjugates of the right ones. In fact: 
\[
\langle\pm|=\frac{1}{\sqrt{2}}\l(e^{h/2}\langle i|\mp e^{-h/2}\langle i+1|\r).
\]

\begin{figure}
\includegraphics[width=7cm]{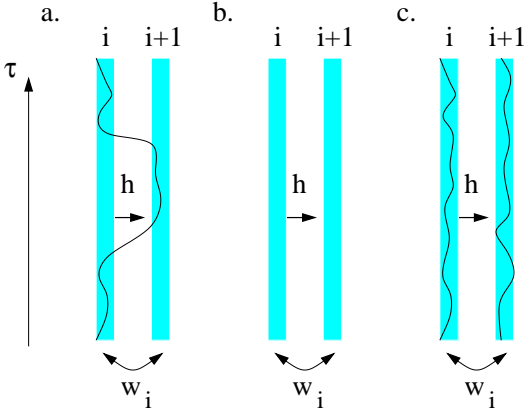}
\caption{Eigenstates of the bond $i$. (a) The lowest eigenstate of the
  Hamiltonian $\H_i$ from Eq. (\ref{2.1}) is $|-\rangle$
  (Eq. \ref{2.5}). 
Sites $i$ and $i+1$ share a single vortex. In the imaginary time
  description, it is as though the vortex tunnels back and forth
  between the two columnar pins. The energy of this state is $-\mu-w_i$.  
  (b) An excited states in which both pins are empty, with energy eigenvalue zero. 
  (c) An excited eigenstate of $\H_i$ with energy $-2\mu$; both pins are occupied by vortices.
    \label{eigenstate}}
\end{figure}

As long as $E_-<0$ and $E_+ > 0$ the ground state of $\H_i$ from Eq. (\ref{2.1})
will contain a single vortex split
between sites $i$ and $i+1$. The eigenstate of bond $i$ is therefore
\be
|-\rangle=\l(e^{-h/2}\bd_i+e^{h/2}\bd_{i+1}\r)|0\rangle.
\label{2.5}
\ee
This state is described in Fig. \ref{eigenstate}a.

The next step is to include the rest of the Hamiltonian, and in
particular neighboring bonds corresponding to  $\H_{i\pm 1}$. These terms will lead to quantum
fluctuations above the ground state of $\H_{I}$. Thus it is necessary
to find the allowed excited states of (\ref{1.1}) which involve 2 and
0 vortices. These states are simply
\be
\ba{c}
|1,1\rangle=\bd_i \bd_{i+1}|0\rangle,\vspace{2mm}\\
|0,0\rangle=|0\rangle,
\ea
\label{2.6}
\ee 
with energies $E=-2\mu$ and $E=0$ respectively; they are
depicted in Fig. \ref{eigenstate}b, and  \ref{eigenstate}c. Note that
their kinetic energy is $E_k=0$; their only energy comes from the chemical potential.

\subsection{Second order perturbation theory and renormalized hopping
  and tilt}

Having diagonalized the bond $\H_i$ we now need to incorporate the rest of the
Hamiltonian as a perturbation. To second order in the hopping matrix elements, we need to concentrate
only on the two neighboring bonds:
\be
\ba{c}
V=-w_{i-1}\l(\bd_{i-1} \b_{i} e^{-h}+\bd_{i} \b_{i-1}e^{h}\r)\vspace{2mm}\\-w_{i+1}\l(\bd_{i+1} \b_{i+2} e^{-h}+\bd_{i+2} \b_{i+1}e^{h}\r)
\label{2.7}
\ea
\ee
The second order Hamiltonian is then:
\be
\ba{c}
V^{(2)}=-\frac{\langle-|V|0,0\rangle\langle0,0|V|-\rangle+\langle-|V|1,1\rangle\langle1,1|V|-\rangle}{w_i}\vspace{2mm}\\
=-\frac{w_{i-1}^2}{w_i}-\frac{w_{i+1}^2}{w_i}\vspace{2mm}\\
-\frac{w_{i-1}w_{i+1}}{w_i}\l(\bd_{i-1} \b_{i+2} e^{-3h}+\bd_{i+2} \b_{i-1}e^{3h}\r),
\ea
\label{2.8}
\ee
where, as in Eq. (\ref{2.6}), $|0,0\rangle$ is a state with both sites $i$ and $i+1$ empty,
and $|1,1\rangle$ is a state with both sites full.  Quantum fluctuations thus give rise to a new coupling between sites
$i-1$ and $i+2$. This renormalization process is depicted in
Fig. \ref{RSRG}, and is discussed below.

\begin{figure}
\includegraphics[width=8cm]{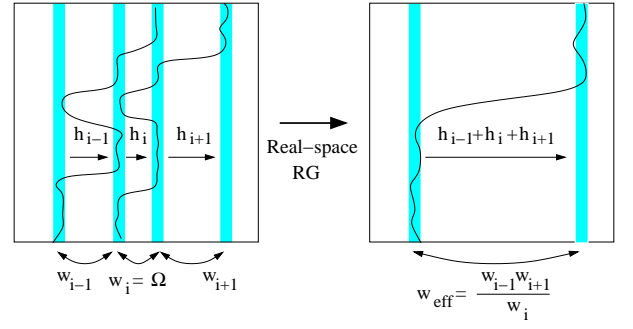}
\caption{Real-space RG decimation step, where we allow for different
  tilt fields $h_j$ on neighboring bonds. In each step of the renormalization we find
  the strongest bond, $w_i=\O$. We diagonalize it and minimize its
  energy by allowing a single vortex to fluctuate between the two
  columnar pins, $i$ and $i+1$. A vortex in site $i-1$ may then
  fluctuate into site $i$, raise the energy of sites $i$ and $i+1$,
  and make the vortex in site $i+1$ relax by fluctuating into site
  $i+2$. Thus we can eliminate sites $i$ and $i+1$, by including
  the effective hopping between sites $i-1$
  and $i+2$. The effective transverse field (tilt) between $i-1$ and
  $i+2$ becomes the sum of the tilts connecting these two sites.
 \lb{RSRG}}
\end{figure}

Eq. (\ref{2.8}) is quite remarkable. It tells us that after localizing
a vortex between sites $i$ and $i+1$, we can forget about these two
sites. Except for a constant contribution, the only change we need to make
to the Hamiltonian is to add a new hopping term that allows vortices
to hop over the occupied pair, from $i-1$ to  $i+2$. In addition, we
see that this hopping is associated with the composite tilt of the
three bonds linking $i-1$ and $i+2$. In general, upon allowing for
different tilt fields on neighboring bonds, the effective strength and
effective tilt of the renormalized bond are respectively  
\be
\ba{c}
w^{eff}_{i-1,\,i+2}=\frac{w_{i-1}w_{i+1}}{\O}\vspace{2mm}\\
h^{eff}_{i-1,\,i+2}=h_{i-1}+h_i+h_{i+1}
\ea
\label{2.9}
\ee
where we introduced the notation $\O=\max\{w_i\}$ (also see Fig. \ref{RSRG}). 
Eqs. (\ref{2.9}) are the {\it RG rules} for the RSRG as applied to the
vortex hopping problem. 

As it turns out, Eqs. (\ref{2.9}) coincide with the RG flow
equations for the fermion hopping problem (or XX spin chain problem) 
in the Hermitian case $h=0$. 
The only difference is that the tilt $h_i$ replaces the length of
bonds, as one might expect in the case of localized states. \cite{HatanoNelson1997}
In order to solve the flow equations we can thus use techniques known 
from the Hermitian case. 

Two important points should be made here:  
(1) The flows of the $h_i$'s and $w_i$'s are independent. (2) No new terms are
produced. In particular, we do not produce randomness in the chemical
potential. In the pure model it is the particle-hole symmetry that
prevented us from having new chemical-potential terms. But Eq. (\ref{1.1}) with nonzero
tilt doesn't have a particle-hole symmetry, or more precisely, it is
symmetric under a particle-hole transformation accompanied by a
changing the tilt direction:
\be
\ba{ccc}
\b\rightarrow\bd, & \bd\rightarrow\b & h\rightarrow -h
\ea
\label{2.10} 
\ee
Terms like 
\[
\bd\b e^h+\b\bd e^{-h}
\]
are allowed in the presence of the modified p-h symmetry
(Eq. \ref{2.10}), and they would create a bias towards vortices (or
holes). Fortunately, as discussed above, they are not produced.


\subsection{Flow equations for the distribution of $h_i$ and $w_i$} 

In the previous section we derived the RG rules that govern the
effective tilt and hopping matrix elements. By using these RG rules 
we can gradually decrease the energy scale of the Hamiltonian while determining the high energy parts of the 
ground state. When the energy scale reaches zero, the Hamiltonian that we started with
is effectively diagonalized. The RG rules on their own, however, do not tell us much about the
eigenvalue spectrum. To obtain useful information, we need to convert the RG rules into differential flow equations
for the distributions of bond strength and tilt. As pointed out
before, this has been essentially done already in 
the corresponding Hermitian problem.

Eq.~(\ref{2.9}) provides RG rules that are identical to the case of
the random XX Hamiltonian. \cite{DSF94} In this case we renormalize the bond strength
and the {\it length} of bonds. Eqs. (\ref{2.9})
suggest that for random vortex hopping in the non-Hermitian problem,
the length of a bond is replaced by the bond's {\it tilt}. 
This observation allows us to use essentially all
results derived for the XX model in Ref. \onlinecite{DSF94} and apply
them in this problem. In this section we review these results as they
apply to the vortex hopping problem. 

We need the flow equation for the joint probability
distribution $\rho(w,\,h)$ of the hopping strength and tilt on a bond 
with respect to the gradually decreasing energy scale $\O$. 
As in the theory of random spin chains, we introduce a logarithmic variable $\zeta$ to replace the bond's
strength, and a renormalization-group flow parameter $\Gamma$:
\be
\ba{c}
\zeta_i=\ln\frac{\O}{w_i},\vspace{2mm}\\
\G=\ln\frac{\O_0}{\O},
\ea
\label{3.1}
\ee
where $\O_0$ is the initial energy scale (largest matrix element) in the
problem. The logarithmic energy scale $\G$ is the RG flow parameter,
and it increases as the energy scale is decreased. Note also that in
every step of the RG we renormalize away bonds with $\zeta\rightarrow
0$. In the continuum limit, one can verify that to generate the flow
$\G\rightarrow\G+d\G$ we renormalize  all bonds with $0\le \zeta<
d\G$.  The logic behind the definition of $\zeta$ in Eq. (\ref{3.1}) becomes clear upon considering the
first RG rule in terms of the new variables. 
Eq.~(\ref{2.9}) takes the form 
\be
\zeta^{eff}_{i-1,i+2}=\zeta_{i-1}+\zeta_{i+1}.
\label{3.2}
\ee

Upon transforming $\rho(w,\,h)$ into the new variables, we obtain the joint probability distribution
$P(\zeta,\,h)$. The flow equation for $P(\zeta,\,h)$ is given by\cite{DSF94} 
\begin{widetext}
\be
\frac{dP(\zeta,h)}{d\G}=\der{P(\zeta,h)}{\zeta}+\int\int
d\zeta_1d\zeta_2 dh_0 dh_1 dh_2
\delta(\zeta-\zeta_1-\zeta_2)\delta(h-h_0-h_1-h_2)
P(0,h_0)P(\zeta_1,h_1)P(\zeta_2,h_2).
\label{3.3}
\ee
\end{widetext}
Eq, (\ref{3.3}) greatly simplifies if we take the Laplace transform
with respect to $h$. If $y$ is the Laplace transform variable conjugate
to $h$, we obtain
\be
\ba{c}
\frac{dP(\zeta,y)}{d\G}=\der{P(\zeta,y)}{\zeta}\vspace{2mm}\\
+\int\int
d\zeta_1d\zeta_2 
\delta(\zeta-\zeta_1-\zeta_2)
P(0,y)P(\zeta_1,y)P(\zeta_2,y).
\label{3.4}
\ea
\ee
The derivation of this equation is given in detail in Ref. \onlinecite{DSF94}.

\subsection{Universal fixed point distributions of $w$ and $h$, 
  length energy scaling, and density of states}

Remarkably, Eq. (\ref{3.4}) has a scaling solution which is an attractor to
essentially all boundary conditions.\cite{DSF95} This solution is 
\be
P(\zeta,\,y)=\frac{1}{\G+\G_0}\exp\l(-\sqrt{y}\coth\l(\sqrt{y}\l(\G+\G_0\r)\zeta\r)\r)
\label{3.5}
\ee
where $\G_0$ is a non-universal constant of integration. This solution, once inverse-Laplace transformed, gives the joint probability distribution
of bond strengths and effective tilt. Given some initial Hamiltonian
and tilt, at sufficiently low energy scales (i.e. in the limit of
large system size for the vortex problem) the distribution of effective couplings will be given by Eq. (\ref{3.5}).

From this solution we find that the average logarithmic coupling obeys:
\be
\overline{\zeta} =\G+\G_0,
\label{3.6}
\ee
where over-bar denotes disorder averaging. Also, Eq. (\ref{3.6}) is obtained from the marginal distribution for the logarithmic
couplings, i.e., Eq. (\ref{3.5}) with $y\rightarrow 0$. 
In addition we infer that the Laplace parameter conjugate to the tilt scales as
\[
\overline{y}\sim (\G+\G_0)^{-2},
\]
and therefore the tilt field scales as
\be
\overline{h_{eff}}\sim (\G+\G_0)^2.
\label{3.7}
\ee
From the above scaling we can infer the average length of  a
bond after the RG reached the logarithmic energy scale $\G$. 
This scaling is identical to the 
tilt-energy scaling, as mentioned above:
\be
l=l_0 (\G+\G_0)^2.
\label{3.71}
\ee
The concomitant relation to the length-energy scaling, is the relation
between the fraction
of unoccupied pinning sites, $\den$, and the energy. This
relation has to be
\be
\den=\frac{\den_0}{\l(\G+\G_0\r)^2}. 
\label{3.72}
\ee
where $\G_0$ and $\den_0$ are non-universal constants that depend on
the particular realization of the disorder. 

A formula used by Shnerb and Nelson,\cite{ShnerbNelson1998,Dahmen} and
Brouwer et al. \cite{Brouwer} will provide us with a
connection between the spectrum of the Hermitian Hamiltonian  (with
zero tilt) and the spectrum of the tilted Hamiltonian, see Sec. \ref{NHF}.  In order to use
this formula to find information about delocalized vortex states,
however, we will need knowledge of the density of states in the
zero-tilt limit. 

The density of states (DOS) can also be derived from
Eq. (\ref{3.72}). The number of states per length in the energy
interval $\Omega\to \Omega-d\Omega$, or $\G\to
\G+d\G$ (with $\Omega=\Omega_I e^{-\G}$) is just the number of bonds
that get decimated at that energy scale. This, in turn, is given by:
\[
\den\cdot P(\zeta=0,y=0)d\G.
\]
Converting this into energy terms, using Eqs. (\ref{3.5}) and
(\ref{3.72}), we obtain 
\be
\f(\epsilon)=\frac{\den_0}{\ln^3\l(\Omega_I/\epsilon\r)}\frac{1}{\epsilon}
\label{3.8}
\ee
where the energy $\epsilon$ is measured from the center of the band,
and 
\be
\O_I=\O_0 e^{\G_0}.
\lb{3.85}
\ee
We expect Eq. (\ref{3.8}) to be asymptotically exact for small
$\epsilon$.

In order to check Eqs. (\ref{3.6}) and (\ref{3.72}), we carried
out the real-space RG procedure numerically on chains with
$5\times 10^6$ sites for two types of distributions - block
distribution and displaced power law distribution. 
In Figs. \ref{numfig1} and \ref{poissonfig1} we
plot the average logarithmic coupling and the density of free sites vs. the logarithmic flow
parameter, $\G$, for the two distributions types (further described in
the figure captions). From these plots we also infer the values of the
non-universal constants $\den_0$ and $\G_0$, which will be used later to
compare results for the vortex localization problem.  

\begin{figure*}
\includegraphics[width=11cm]{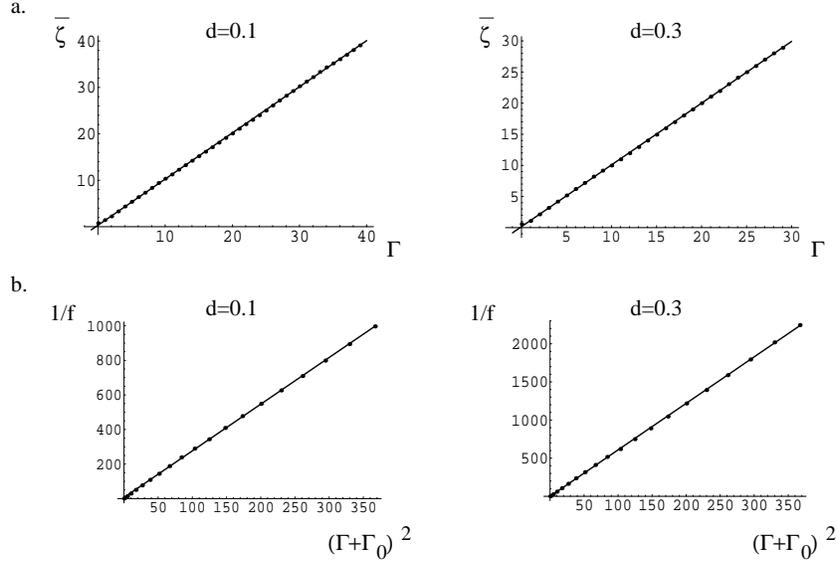}
\caption{
Numerical real-space RG for an initial block distribution, i.e.,
 initial hoppings uniformly distributed between $d$ and $1+d$. Each graph
 represents averaging over 20 realizations. (a) The average logarithmic
 coupling, $\overline{\zeta}$ vs. the logarithmic RG flow parameter, $\G$. The
 fitted slope in the two curves is 0.99. From the intercept we deduce
 $\G_0$ for the two curves: $\G_0=0.31$ ($d=0.1$) and  $\G_0=0.19$
 ($d=0.3$). These plots should be compared with the linear
 dependencies predicted by Eq. (\ref{3.6}). (b)
 The inverse density of free sites $1/\den$ vs. $(\G+\G_0)^2$, with $\G_0$ obtained from the
 graphs in (a). As expected from Eq. (\ref{3.72}), the intercepts of the
 curves with the y axis in the
 two plots are negligible compared to $1/\den$: $2.6$ ($d=0.1$) and
 $-2.4$ ($d=0.3$). From the slope of the two curves we find:
 $\den_0=0.37$ ($d=0.1$) and $\den_0=0.16$ ($d=0.3$). 
 \lb{numfig1}}
\end{figure*}

\begin{figure*}
\includegraphics[width=11cm]{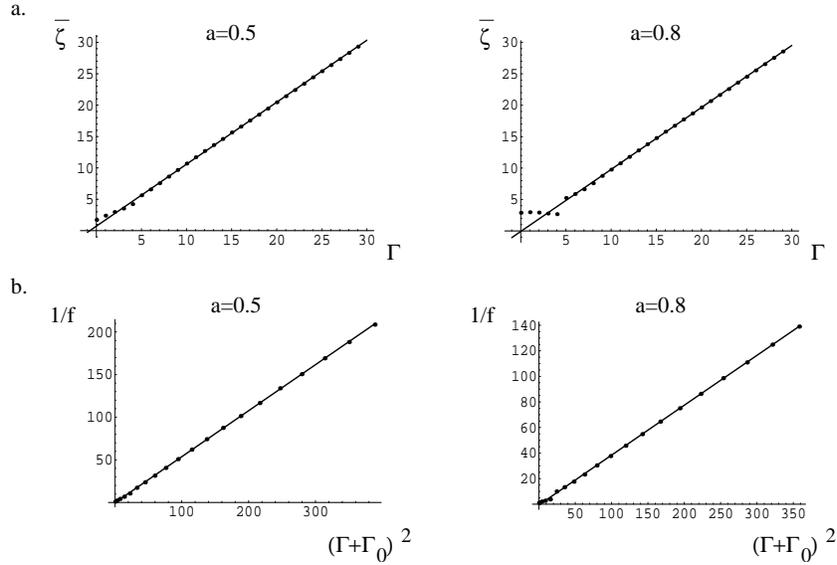}
\caption{
Numerical real-space RG for an initial 'displaced power-law' distribution, with
 initial hoppings given by $w=0.01+w_r$ and $0<w_r\le 1$ distributed
 as $P[w_r]=(1-a)\frac{1}{w_r^a}$. The distribution of $w_r$
 corresponds to a hopping strength which is exponentially suppressed with
 the distance between pins: $w_r\propto e^{-x}$ with $x$ distributed as
 $P[x]\propto e^{-(1-a)x}$. Each graph
 represents averaging over 20 realizations. (a) The average logarithmic
 coupling  $\overline{\zeta}$ vs. the logarithmic RG flow parameter  $\Gamma$. The
 fitted slope in the two curves is 0.99. From the intercept we deduce
 $\G_0$ for the two curves: $\G_0=0.71$ ($a=0.5$) and  $\G_0=-0.08$
 ($a=0.8$). As before, these plots should be compared with Eq. (\ref{3.6}). (b)
 The inverse density of free sites $1/\den$ vs. $(\G+\G_0)^2$, with $\G_0$ obtained from the
 graphs in a. As expected from Eq. (\ref{3.72}), the intercepts of the
 curves with the y axis in the
 two plots are negligible compared to $1/\den$: $-0.29$ ($a=0.5$) and
 $-0.6$ ($a=0.8$). From the slope of the two curves we find:
 $\den_0=1.85$ ($a=0.5$) and $\den_0=2.6$ ($a=0.8$). 
 \lb{poissonfig1}}
\end{figure*}

\subsection{Applicability of the RSRG with nonzero tilt}

The application of the RSRG breaks down as soon as we reach energies
at the mobility edge, because we can no longer write a
vortex wave function which is localized between two sites - this
becomes impossible as soon as the chemical potential reaches the
mobility edge, and the states become extended. We {\it can} nevertheless take
advantage of the real-space RG as long as we stop decimating when the
energy scale reaches the mobility edge. In addition, the RSRG gives a reliable expression for the
density-of-states of the Hamiltonian Eq. (\ref{1.1}) with zero-tilt.

\section{Connection with vortex physics\lb{conn}}

In the previous section we derived the RSRG for the vortex-hopping imaginary time
Hamiltonian. In this section we will clarify the relation of the
results obtained above to the physics of vortex lines with columnar
pins in planar superconductors. 

The vortex hopping Hamiltonian yields an energy spectrum, with 
filling factor determined by a chemical potential,
$\mu$. The RSRG outlined above is used to determine the wave function
of vortices in a given chemical potential; in each step of the RG, the
strongest bond is diagonalized, and if the bond-energy of a vortex placed
on the bond is lower than the chemical potential,
\be
E_-=-\tilde{w}-\mu<0,
\label{3.81}
\ee
then a vortex is localized on that bond. 
If the strongest bond in the renormalized chain is too weak to obey
(\ref{3.81}), then the RG is stopped and we have found the ground state of
the zero-tilt model. 

For flux lines, however, we often want to know the longitudinal
magnetic field, proportional to the vortex density $\nn$. We need to find out how $\nn$ and $\mu$ are
related. This is done using Eq. (\ref{3.72}). The longitudinal vortex density
vs. the fraction of unoccupied sites $\den$ is
\be
\nn=\frac{1}{2}(1-\den)=\frac{1}{2}\l(1-\frac{\den_0}{\l(\G+\G_0\r)^2}\r),
\label{3.82}
\ee
i.e. the longitudinal vortex density is half the density of filled pinning
sites, since each vortex is localized on bonds, and thus on two sites. The number of unoccupied sites $\den$ as a function of energy scale is
taken from Eq. (\ref{3.72}). Now, the condition (\ref{3.81}) gets
translated to (assuming negative $\mu$; for positive $\mu$ we can
carry out a particle-hole transformation)
\be
\G=\ln\frac{\O_0}{|\mu|}.
\ee
Upon inverting Eq. (\ref{3.82}), we thus obtain
\be
\mu=-\O_0e^{\G_0}\exp\l(-\sqrt{\frac{\den_0}{1-2\nn}}\r).
\label{3.83}
\ee
Note that $\O_0$ is the initial energy scale, and $\den_0$ and $\G_0$ are constants that are
determined by the initial distribution of hoppings, and which can be calculated as
in Fig. \ref{numfig1}. 
Eq. (\ref{3.83}) is tested in Figs. \ref{numfig2} and \ref{poissonfig2} for two
forms of disorder. 

\begin{figure}
\includegraphics[width=8cm]{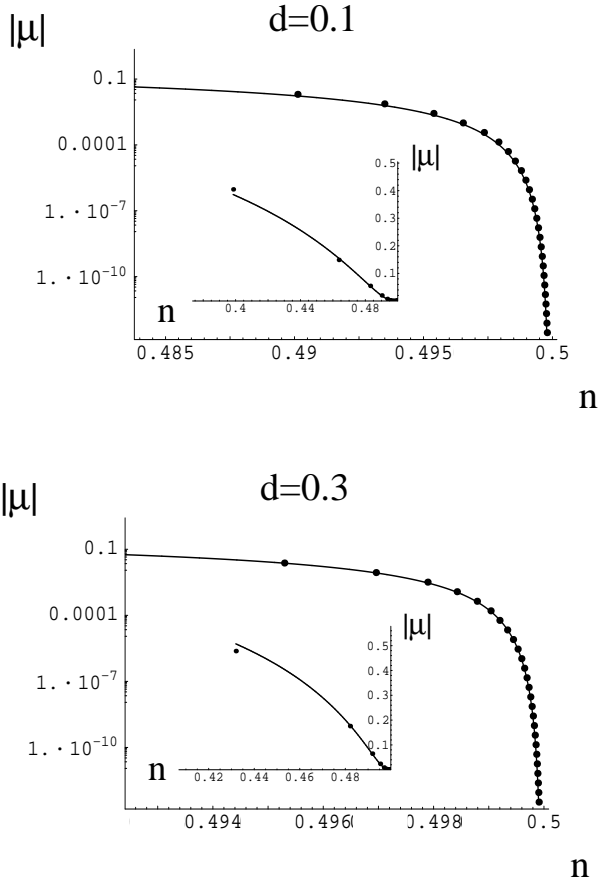}
\caption{Semi-log plots of the vortex chemical potential, $\mu$, vs. the longitudinal vortex density,
  $n$, for a box disorder distribution and two values of $d$. The solid lines are
  Eq. (\ref{3.83}) with $\G_0$ and $\den_0$ given in the caption of
  Fig. \ref{numfig1}. Insets show the linear-linear plots.
   \lb{numfig2}}
\end{figure}

\begin{figure}
\includegraphics[width=8cm]{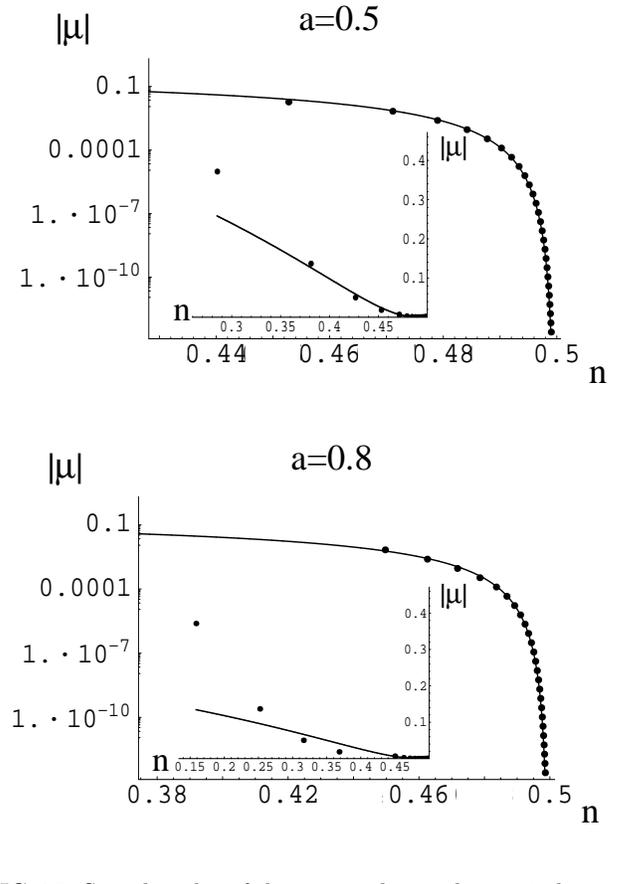}
\caption{Semi-log plot of the vortex chemical potential, $\mu$, vs. the longitudinal vortex density,
  $\nn$, for two displaced power-law distributions as in Fig. \ref{poissonfig1}. The solid lines are
  Eq. (\ref{3.83}) with $\G_0$ and $\den_0$ given in the caption of
  Fig. \ref{poissonfig1}.  Insets show the linear-linear plots.
   \lb{poissonfig2}}
\end{figure}

In the following, we will find the mobility edge, $\mu_m$, of the
vortex-hopping model with tilt, using the RSRG. We will then use the
above relations to convert the results to the physical variable
$n$. 

All results will depend on the initial conditions
through the non-universal quantities $\O_I$ and
$\den_0$. Nevertheless these results are valuable since our results will
be able to describe the entire phenomenology of the depinning
transition in terms of these two constants, which encode the relevant
part of the initial distribution. These constants can be evaluated for
a given distribution either by carrying out the RSRG numerically, or
by fitting curves to partial results obtained in other methods, for
instance, by comparing the density of states, Eq .(\ref{3.8}), to a
spectrum obtained numerically. 

In general, when comparing our results to spectra obtained
numerically, it is easier to use formulas with $\mu$ in them rather
than the physical $\nn$. 


\section{Application of the RSRG results to non-Hermitian delocalization \lb{results}}

In the above we derived many characteristics of the vortex hopping
problem while assuming that all vortices are localized. The most
interesting aspects of the problem, however, arise in relation to the
{\it delocalization transition} of vortices with sufficient tilt. 

The physical quantities of interest are essentially all related to the
energy-spectrum of the problem. Most particularly, given a tilt - an
angle relative to the columnar defects with which we apply a magnetic
field - how much magnetic field can we apply such that all vortex
lines remain localized in configurations parallel to the columnar pins? This question is answered by finding the
mobility edge of the non-Hermitian boson hopping model. Once we know
the effective chemical potential in which the vortex bosons become depinned, the
density of bosons at this chemical potential determines the critical
field for destruction of the transverse Meissner effect. 

Another interesting quantity is the transverse magnetic field as a
function of tilt above the mobility threshold for a given
longitudinal field. The transverse field is proportional to the boson
current [see Eq. (\ref{current_op0})]. 
For tilts slightly above the
mobility threshold $h_c$ (i.e., a chemical potential slightly above the mobility
edge), the boson current is simply given as (Eq. \ref{7.8}):
\[
J=\frac{1}{\pi}(\mu-\mu_c)\tan \theta
\] 
where $\theta$ is the angle of ascent of the spectrum in the complex
energy plane (see Fig. \ref{fig3}). The angle $\theta$ is a universal function
of the external magnetic field, or the chemical potential. 

In this section we will analytically derive these quantities for the
random hopping model. First, we
will find the mobility edge as a function of tilt, or, equivalently, the tilt threshold
as a function of external field. In order to obtain the transverse
magnetization, we will need to use a formula appearing in
Refs. \onlinecite{Brouwer} and
\onlinecite{ShnerbNelson1998,Dahmen} , which connects the spectrum of the
Hermitian boson hopping Hamiltonian with the spectrum of the
non-Hermitian problem and the tilt. We will introduce this formula,
and use it to derive the quantities of interest. 

\subsection{Mobility edge of the random hopping Hamiltonian \label{perp}}

By using the RSRG, we demonstrated in Sec. \ref{RSRGsec} that when the
tilt is zero, vortices are localized between pairs of pinning sites. We
also showed that as we reduce the energy scale of the Hamiltonian, and
place vortices onto pairs of pinning sites that are increasingly far
apart, the effective tilt, $h_{\rm eff}$ per site also increases. For any
given initial tilt, at some energy scale there will no longer be a
localized vortex solution. This energy is the mobility edge,
$\mu(h)$. Alternatively, there will be a
critical longitudinal vortex density, which we denote $n(h)$, above
which vortices are no longer restricted to the pinning sites. In this
section we find the relation between the tilt and the critical
longitudinal vortex density. 

\subsubsection{Single strong bond}

\begin{figure}
\includegraphics[width=8.5cm]{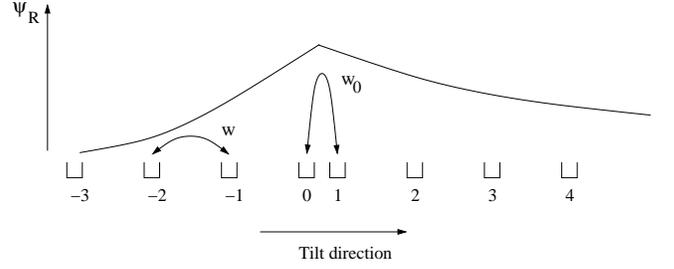}
\caption{Simplified model with a single strong pin. We use this model to
  determine the mobility edge for a given tilt. The model consists of
  a uniform pinning lattice with hopping $w$, periodic boundary conditions, 
  and one pair of strongly interacting sites with hopping $w_0$. The strong bond localizes a
  vortex in it as long as the tilt is sufficiently weak. The wave
  function is schematically plotted, showing that the its decay is
  asymmetric. Strong enough tilt will prevent the wave function from
  decaying on its right side, signaling a delocalization
  transition. This problem with non-hard-core vortices was
  considered in Ref. \onlinecite{Halperin}. \label{fig2}}
\end{figure}

We start our investigation by solving a simplified problem that will
give us the correct solution. We will then proceed to justify it for
the random hopping Hamiltonian. Consider  a ring of columnar defects,
where there is one strong
bond, $w_0$, and all other bonds are $w\ll w_0$ (Fig. \ref{fig2}):
\be  
\ba{c}
\H=-\summ_{i\neq 0} w\l(\bd_i \b_{i+1} e^{-h}+\bd_{i+1} \b_{i}
e^{h}\r)\vspace{2mm}\\
-w_0\l(\bd_0 \b_{1} e^{-h}+\bd_{1} \b_{0}
e^{h}\r).
\label{4.1}
\ea
\ee
The strong bond can localize a single vortex; we will determine at
what tilt $h$ this state delocalizes.

The eigenfunctions of Hamiltonian (\ref{4.1}) can be solved
exactly. In particular, the localized wave function at the bottom of
the band reads, up to
normalization,
\be
\ket{\psi}=\l(\summ_{n\leq 0}e^{\kappa_L \l(n-1/2\r)}\bd_n+\summ_{n>0}e^{-\kappa_R \l(n-1/2\r)}\bd_n\r)\ket{0}.
\label{4.2}
\ee
The left and right decay coefficients, $\kappa_L$ and $\kappa_R$, are given by:
\be
\ba{c}
\kappa_R a=\ln\frac{w_0}{w}-h,\vspace{2mm}\\
\kappa_L a=\ln\frac{w_0}{w}+h,\vspace{2mm}\\
\ea
\label{4.3}
\ee
where $a$ is the lattice constant.

A localized solution of the form Eq. (\ref{4.2}) can only exist as
long as both $\kappa_R$ and $\kappa_L$ are positive; otherwise we cannot
accommodate the periodic boundary conditions (as the system is a
ring). Hence, the threshold tilt is given by the condition that
$\kappa_R=0$, or
\be
h_c=\ln\frac{w_0}{w}.
\label{4.4}
\ee

\subsubsection{Generalization for the random-hopping model}

Let us now generalize this result to the case of random
hopping. At every stage of the RG we assume that the strongest bond localizes
a vortex, with energy $E=\O=\Omega_0 e^{-\G}$. If we assume that all
other bonds are the same, the problem reduces to the localizing bond
problem above. However, we need a
generalized condition that takes into account the randomness of the
weak bonds. In this case the natural generalization of condition
(\ref{4.4}) would be:
\be
h^c_{\rm eff} =\overline{\ln\frac{w_0}{w}}=\overline{\ln\frac{\O}{w}}=\G+\G_0.
\label{4.5}
\ee
where the last equality is due to Eq. (\ref{3.6}). The reason for writing $h_{\rm eff}$ is to remind 
ourselves that the tilt $h$ we consider in this section is the renormalized $h$. In the next section
we connect the results derived above with the bare parameters of the
vortex-pinning problem. 

So far Eq. (\ref{4.5}) is just a physically motivated guess. There are, however,
several more rigorous ways to derive it. The most straightforward
proof is obtained via perturbation
theory. Consider the Hamiltonian Eq. (\ref{2.1}), and assume that
coupling $w_0$ is by far the strongest in the chain. From
Eq. (\ref{2.4}) we know what the zeroth order solution for the wave
function is:
\be
|-\rangle=\frac{1}{\sqrt{2}}\l(e^{-h/2}\bd_0+ e^{h/2}\bd_1\r)\ket{0}
\label{4.6}
\ee
From perturbation theory, we can easily see that the wave function at
site $\ket{m}$, to lowest non-vanishing order in the $w_i$'s reads (assume $m>0$) 
\begin{widetext}
\be
\langle m|-\rangle\approx \l(\prodd_{i=2}^{m}\frac{\langle i|w_i
  e^h\bd_i \b_{i-1}|i-1\rangle}{w_0}\r)\frac{\langle 2|w_{1}
  e^h\bd_{2}\b_{1}|-\rangle}{w_0}=\frac{e^{h/2}}{\sqrt{2}}\prodd_{i=1}^m \l(\frac{e^{h}w_i}{w_0}\r)^{m-1}
\label{4.7}
\ee
\end{widetext}
By requiring that the product at the end of Eq. (\ref{4.7}) remains
finite as $m\to \infty$,  we obtain condition (\ref{4.5}).
Another derivation is supplied in Appendix \ref{appA}. 


\subsubsection{Critical tilt and field}

As explained in Sec. (\ref{RSRGsec}) above, the RSRG eliminates the strongest hoppings in the Hamiltonian
(\ref{1.1}) and gradually populates localized vortex states, until the
energy scale of the renormalized Hamiltonian reaches the chemical
potential:
\[
\O=\max_i\{\tilde{w}_i\}=-\mu,
\]
where the tilde indicates that the maximum is evaluated over the set
of remaining (and renormalized) bond strengths. The chemical
potential, $\mu$, in this case is a tuning parameter that controls the vortex
filling factor of the lattice, and hence the longitudinal vortex density.
In terms of the RG variables, $\Omega=\O_0e^{-\G}$, and therefore we can
define
\be
 \G_{\mu}=\ln\frac{\O_0}{-\mu}.
\label{5.1}
\ee

From the RG procedure, and in particular from Eq. (\ref{3.7}), we know that 
\be
\overline{h}_{\rm eff}\approx  h_0\l(\G_{\mu}+\G_0\r)^2/\den_0.
\label{5.2}
\ee
where $\G_0$ and $\den_0$ were defined in Eqs. (\ref{3.6}) and (\ref{3.72})
respectively, and $h_0$ is the unrenormalized tilt per lattice site.

In order to find a relationship between the mobility edge of the
vortex hopping problem and the unrenormalized tilt $h$, we substitute the relation 
$h_{eff}^c=h_c\l(\G_{\mu}+\G_0\r)^2/\den_0$ into Eq. (\ref{4.5})
and obtain 
\be
h_c=\frac{\den_0}{\G_{\mu}+\G_0}.
\label{5.22}
\ee

The physical quantity in the vortex pinning problem that the parameters
$\mu$ and $\G_{\mu}$ determine is the vortex density, $\nn$. As a function of $\G_{\mu}$, we
showed in Sec. (\ref{conn}) that 
\[
\nn=\frac{1}{2}-\frac{\den_0}{2 \l(\G_{\mu}+\G_0\r)^2}
\]
and therefore the connection between the critical tilt and the
longitudinal field is
\be
h_c=\sqrt{\den_0\l(1-2\nn\r)} 
\label{5.23}
\ee
This is the main result of this section. At a given field $\nn$, as we
increase  the bare tilt, when the tilt reaches $h_c$, vortices get
delocalized and transverse flux ensues. The critical tilt
vanishes as $\nn\to 0.5$, i.e., as vortex states fill the lattice upto
the middle of the band. This is since the localization length of the
tilt-free vortex states diverges as the chemical potential approaches
zero.   
We demonstrate the validity of Eq. (\ref{5.23}) using the numerical
real-space RG in Figs. \ref{numfig3} and \ref{poissonfig3}. 
Note that the RSRG approach is most accurate near the middle of the
band, where $\nn\sim 0.5$, in the region where the density of states
diverges as in Eq. (\ref{3.8}), also see Sec. \ref{numerics}.

\begin{figure}
\includegraphics[width=7cm]{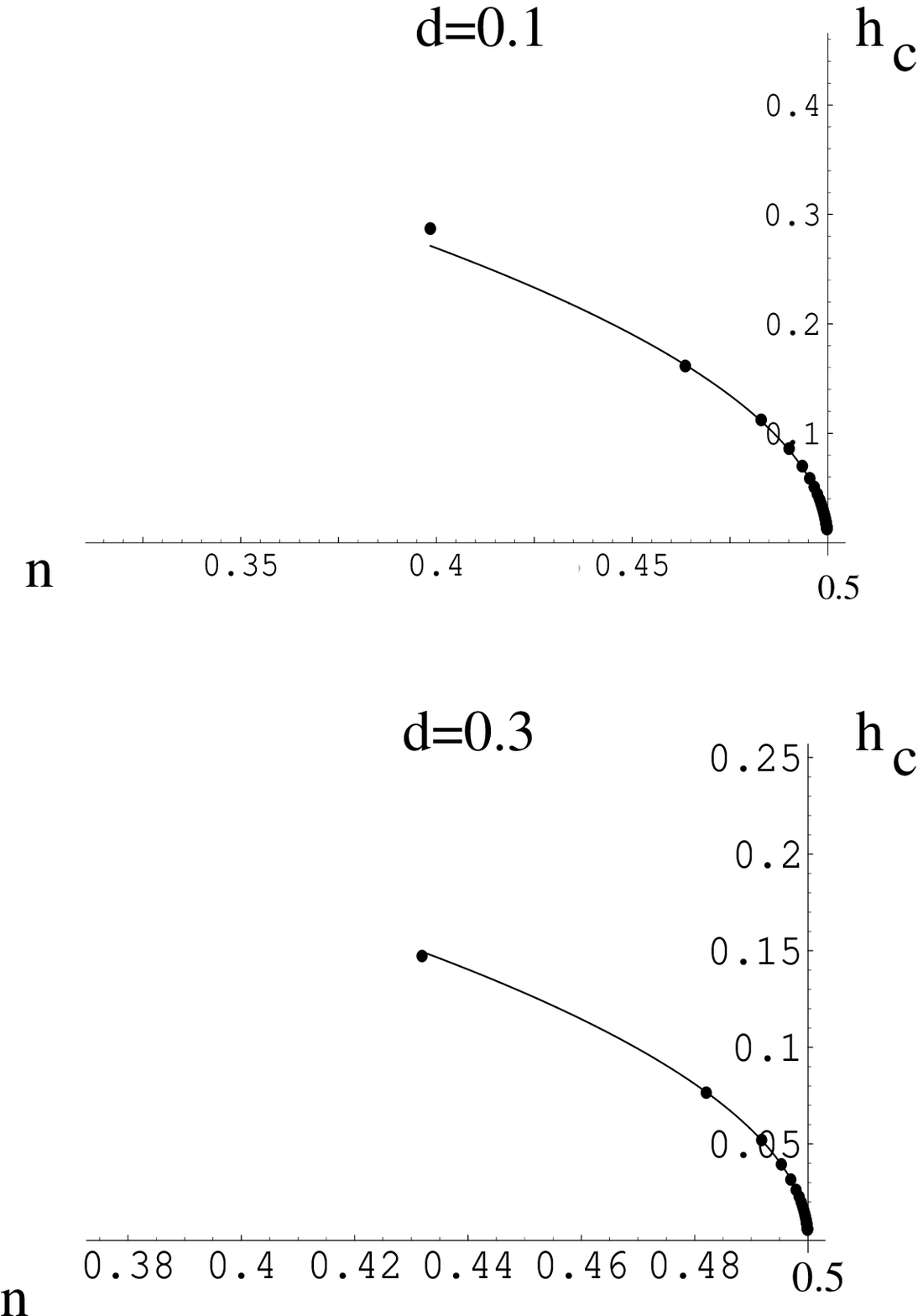}
\caption{Critical tilt, $h_c$, vs. longitudinal vortex density,
  $\nn$, for the two box-distributions from Fig. \ref{numfig1}a. The dots are the result of the numerical implementation of the
  RG, where the critical tilt is the sum of $\zeta$ divided by the
  full (bare) length of the chain. The solid line is a plot of
  Eq. (\ref{5.23}).
 \lb{numfig3}}
\end{figure}

\begin{figure}
\includegraphics[width=7cm]{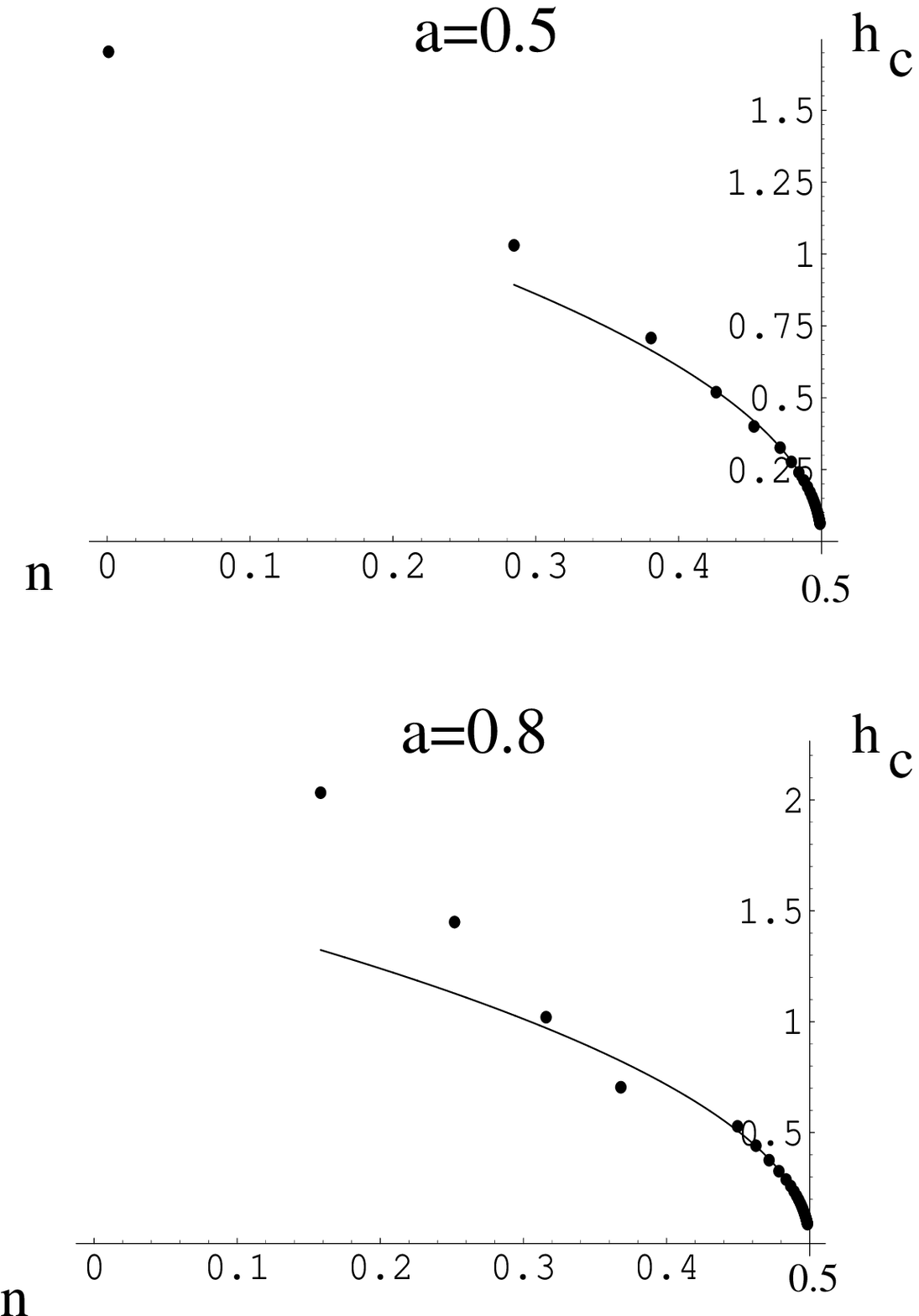}
\caption{Critical tilt, $h_c$, vs. longitudinal vortex density,
  $\nn$, for the two displaced power-law distributions from Fig. \ref{poissonfig1}. The dots are the result of the numerical implementation of the
  RG, where the critical tilt is the sum of $\zeta$ divided by the
  full (bare) length of the chain. The solid line is a plot of
  Eq. (\ref{5.23}).
\lb{poissonfig3}}
\end{figure}

\subsubsection{Vortex localization length}

Another quantity of interest is the vortex localization length. 
According to
Ref. \onlinecite{HatanoNelson1997}, this length, $\xi_{\perp}$, is
defined as the decay distance of the ket
describing the least localized vortex in the tilt direction. From Eq. (\ref{4.3}) and its
generalization according to Eq. (\ref{4.5}) we see that
$\tilde{\xi}_{\perp}=\frac{1}{\kappa_R}\approx
\frac{1}{\overline{\ln\frac{\O}{w}}-h_{eff}}$, where the tilde
indicates that the length is a renormalized length, and not bare length.
It follows that
\be
\tilde{\xi}_{\perp}=\frac{h_c}{\G_{\mu}+\G_0}\frac{1}{h_c-h}
\label{5.4}
\ee
To convert the renormalized length to physical length, we need to
divide by the fraction of empty pinning sites, given by
Eq. (\ref{3.72}). In terms of physical (bare) length scales, the localization length is 
\be
\xi_{\perp}=\frac{1}{h_c-h}.
\label{5.5}
\ee


\subsection{Non-Hermitian spectral formula \lb{NHF}}

A formula in Ref. \onlinecite{ShnerbNelson1998,Dahmen} (see also
Ref. \onlinecite{Brouwer}) relates the
imaginary part of the eigenenergies of delocalized states with the
spectrum of the Hermitian hopping problem, in the absence of a
tilt. This connection, along with our detailed knowledge of the
random-singlet phase allows us to probe even the delocalized states,
even though the RSRG does not strictly apply for these states. This remarkable
state of affairs stems from the analytic properties for a next-neighbor
hopping problem. 

The non-Hermitian spectral formula is discussed extensively in Ref.
\onlinecite{ShnerbNelson1998,Dahmen}, but we quote it here for completeness:
\be
\prodd_{i=1}^{N}\l(E-\epsilon_i\r)=2\l[\cosh(hN)-1\r]\prodd_{i=1}^{N} \l(-w_i\r),
\label{6.1}
\ee
where $E$ is an eigenvalue of the non-Hermitian vortex hopping problem with
tilt $h$ per site. There are a total of $N$ sites arranged in a
ring. The $\epsilon_i$'s are the eigenvalues of the zero-tilt Hermitian problem.
This is a complex equation, and we will use it primarily to find a
relation between the imaginary part of $E$ and the the excess tilt
$h-h_c$. In principle, we can use the above formula to find {\it all}
eigenvalues of the non-Hermitian problem, since we know the spectrum
of the zero-tilt problem from the RSRG. This strategy can be implemented
numerically.

A particularly useful form of the non-Hermitian spectral formula is as
follows: Let us assume that $N$ and $Nh$ are large, and take the logarithm of Eq. (\ref{6.1}). We obtain 
\be
\summ_{i=1}^{N}\ln\l(E-\epsilon_i\r)=N|h|+ \summ_{i=1}^{N}
\ln\l(w_i\r)+i\pi N.
\label{6.2}
\ee

In Appendix \ref{appA} we show how the RSRG result Eq. (\ref{4.5}) can be deduced from the
discrete non-Hermitian spectral formula. There it is also shown that
we can, in fact, apply the RSRG directly to formula (\ref{6.2}). The
advantage of doing so is that in advanced stages of the RSRG the
distribution of $w_i$ is known from the fixed point solution,
Eq. (\ref{3.5}).

An important formulation of Eq. (\ref{6.2}) is obtained by taking its
real part:
\be
\summ_{i=1}^{N}\frac{1}{2}\ln\l(({\rm Re} E-\epsilon_i)^2+({\rm Im} E)^2\r)=N|h|+ \summ_{i=1}^{N}
\ln\l(w_i\r).
\label{6.25}
\ee
Eq. (\ref{6.25}) specifies a curve in the complex E-plane on which the
eigenvalues of the vortex hopping problem lie. In the absence of the
imaginary part of this equation it is impossible to determine the
location of
individual eigenvalues. But if we are interested in the shape of
the curve, Eq. (\ref{6.25}) is all that is necessary. 

In the following we will make use of the above formulas in the
context of the random hopping problem to obtain analytical expressions
that describe the depinning transition and the spectrum of the
delocalized states.

\subsection{Angle of approach at the mobility edge}

\begin{figure}
\includegraphics[width=8cm]{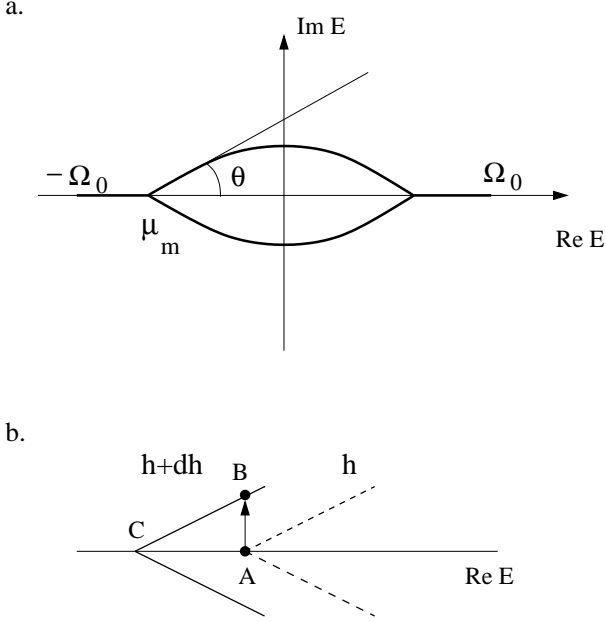}
\caption{(a) Typical spectrum of the random vortex-hopping Hamiltonian with
  non-zero tilt.  The energy eigenvalues lie on the bold lines. Localized states lie on the real axis, while
  delocalized states have an imaginary component. $\O_0$
  is the strongest bond in the bare Hamiltonian, and is of the order
  of $\O_I$, which is the parameter used to define $\G$ - the
  logarithmic energy scale. $\theta$ is the angle of approach of the
  complex branch of the spectrum, and is a critical property of the
  depinning transition. (b) To calculate $\theta$ we concentrate on
  the wedge at the mobility edge. The bold line is the spectrum at
  tilt $h+dh$ and the dashed is the complex spectrum at tilt
  $h$. \label{fig3}}
\end{figure}

An interesting quantity, which enters the critical properties of
the depinning transition, is the angle of approach of the delocalized
branch of the spectrum relative to the real axis, as shown in Fig
\ref{fig3}. This quantity is related to the current of
delocalized 'bosons' (i.e., the number of tilted vortices), as is discussed in the next section. 

The calculation of the angle of approach is somewhat subtle, and we
refer to Fig. \ref{fig3}b to aid the discussion. The figure shows the
line on which the spectrum lies when the tilt is $h+dh$ in bold. The
dashed bold lines indicate the line of spectrum for the lower tilt
$h$. The angle of approach is given by the ratio of the segments:
$\tan\theta= AB/AC$.

We first determine the vertical distance between points A and B in the
figure. For this purpose we differentiate  the real part of the
non-Hermitian spectral formula, Eq.~(\ref{6.25}) with respect to ${\rm Im}
E$ while keeping ${\rm Re} E$ constant. To avoid confusion we use a
representation in terms of differences, taking the difference between
the expression with ${\rm Im} E=c$ and ${\rm Im} E=c+de$. Upon passing
to a continuum representation and introducing a density of states
$\f(\epsilon)$, we obtain
\be
\intt_{-\O_0}^{\O_0}d\epsilon\cdot \f(\epsilon)\frac{c\cdot de}{({\rm Re}
  E-\epsilon_i)^2+c^2}=dh.
\label{7.1}
\ee
In the limit of $dh\to 0$, the integrand on the left is nothing but
$\f(\epsilon)\cdot de\cdot\pi\delta_{({\rm Re} E-\epsilon)}$, and we obtain
(where we drop the ${\rm Re}$ sign)
$\pi\f(E) de=dh$ and hence find that AB in Fig. \ref{fig3}b is
\be
de=dh\cdot \frac{|E|\ln^3\frac{\O_I}{|E|}}{\pi \den_0},
\label{7.2}
\ee 
where $\f(E)$ is taken from Eq. (\ref{3.8}). In Eq. (\ref{7.2}) $E$ designates the energy of the
mobility edge, so we substitute
\[
E=\mu_m(h).
\]

The second part of the calculation focuses on AC. This is the distance
the mobility edge travels as the tilt changes from $h$ to $h+dh$. But
from Eq. (\ref{5.22}) we can find this quantity directly:
\be
\frac{dh}{d\mu_{m}}=\frac{\den_0}{|\mu|\l(\G_{\mu}+\G_0\r)^2}= \frac{\den_0}{|\mu_m|\ln^2\frac{\O_I}{|\mu|}}.
\label{7.3}
\ee

The angle of approach is now readily found:
\be
\tan\theta=\frac{de}{d\mu_m}=\frac{dh}{d\mu_m}\cdot\frac{de}{dh}=\frac{\G+\G_0}{\pi}=\frac{1}{\pi}\ln\frac{\O_I}{|\mu_m|}.
\label{7.4}
\ee
Again it is a universal form, which depends on the initial distribution
only through a single parameter, $\O_I$. Note that as the tilt $h$
approaches zero, and the mobility edge approaches the 'middle of the
band' (i.e., $\mu_m\to 0$), the angle $\theta$ approaches $\pi/2$. 

\subsection{Vortex current near the mobility edge \label{vcurrent}}

The imaginary vortex current near the onset is in general given by
Eq. (\ref{7.8}) in App. \ref{appcurrent}:
\be
J_{total}=2\frac{1}{L}\summ_{k_n}{\rm Re}\der{E_n}{h}=\frac{1}{\pi}\l({\rm Im} E_{\mu}-{\rm Im} E_{\mu_m}\r),
\ee
where $L$ is the number of sites in the lattice. In particular, we can apply this formula to the random hopping problem. In Eq. (\ref{7.4}) the angle of approach near the delocalization transition was found (also see Fig. \ref{fig3}):
\[
\frac{d{\rm Im} E}{d\epsilon}=\tan\theta=\frac{1}{\pi}\ln\frac{\O_I}{|\mu_m|},
\]
where we define $\epsilon$ as:
\[
\epsilon={\rm Re} E,
\]
and $\mu_m$ is the chemical potential at the mobility edge. 
Now, near the onset of transverse flux penetration we can write:
\be
\ba{c}
J_{total}=\frac{1}{\pi}\l({\rm Im} E_{\mu}-{\rm Im} E_{\mu_m}\r)\vspace{2mm}\\
\approx \frac{1}{\pi}\tan\theta(\mu-\mu_m)=\frac{1}{\pi^2}\ln\frac{\O_I}{|\mu_m|}(\mu-\mu_m).
\lb{7.9}
\ea
\ee
This result can also be expressed in terms of the difference between the
longitudinal vortex density and its critical value. The
number of extra vortices per length is just $\nn-\nn_c$. Therefore the
transverse vortex density, i.e., the imaginary 'boson' current is also:
\be
J_{total}\approx \frac{1}{L}{\rm Im}\der{\epsilon}{k} \l(\nn-\nn_c\r)\approx
\l(\nn-\nn_c\r)\frac{\tan\theta}{\pi\cdot\f(\mu_m)}
\lb{7.10}
\ee
Close to $\nn_c$, we may assume that the
real parts of the energy of eigenstates near the mobility edge do not
change as they undergo the delocalization transition. Therefore we can use Eq. (\ref{3.8}) for the DOS.
Upon combining Eq. (\ref{3.8}) with Eqs. (\ref{3.83}), (\ref{5.22}), and (\ref{5.23}) we can write result  (\ref{7.10}) in terms of 
the longitudinal vortex density only: 
\be
J_{total}\approx \l(\nn-\nn_c\r) \cdot\frac{\O_I \den_0}{\pi^2}\frac{1}{(1-2\nn)^2} e^{-\sqrt{\den_0/(1-2\nn)}},
\lb{7.11}
\ee
or in terms of the tilt only: 
\be
J_{total}\approx \l(h-h_c\r) \cdot \frac{\O_I \den_0^2}{\pi^2}\frac{1}{h^3}e^{-\den_0/h}.
\lb{7.12}
\ee
Eqs. (\ref{7.11}) and (\ref{7.12}) are the main results of this section.

One can actually use the above reasoning to obtain not only the total
current, but the current of individual states near the mobility edge. Consider the state $E_{n}$ which can 
be associated with a wave vector $k_n$.
As we change $k=k_n$ to $k_n+\frac{2\pi}{L}$, we move from the
eigenvalue $E_n$ to $E_{n+1}$. Thus 
\be
{\rm Re} (dE)=\epsilon_{n+1}-\epsilon_n\approx\frac{2}{L\cdot\f(\epsilon)}
\lb{7.13}
\ee
where we make use of Eq. (\ref{3.8}), and again set $\epsilon={\rm Re} (E)$.
The factor $2$ in the numerator of Eq. (\ref{7.7}) is due to the real spectrum splitting into two branches, 
and we choose a convention in which both branches of the spectrum contribute to the DOS. Now we can also
substitute $dk=\frac{2\pi}{L}$, and we obtain for the current of this state 
\be
\ba{c}
J_n=\frac{1}{L}{\rm Re}\der{E_n}{h}\vspace{2mm}\\
=\frac{1}{L}{\rm Im}\der{E}{k}\approx
\frac{2}{L\cdot\f(\epsilon)}\frac{1}{2\pi}\tan\theta=\frac{\tan\theta}{L\pi\f(\epsilon)}.
\lb{7.14}
\ea
\ee
where we used $\tan\theta=\frac{d{\rm Im} E}{d\epsilon}$. The result
(\ref{7.14}) can be written in terms of the longitudinal vortex
density, $\nn$, instead of in terms of $\epsilon$:
\be
J_n\approx \frac{\den_0\O_I}{L \pi^2}\frac{1}{(1-2\nn)^2}e^{-\sqrt{\den_0/(1-2\nn)}},
\lb{7.15}
\ee
Eq. (\ref{7.14}) can also be expressed in terms of the tilt $h$, using
Eq. (\ref{5.23}):
\be
J_n\approx \frac{\den_0^3\O_I}{L\pi^2}\frac{1}{h^4}e^{-\den_0/h}.
\lb{7.16}
\ee

\begin{figure}
\includegraphics[width=8.5cm]{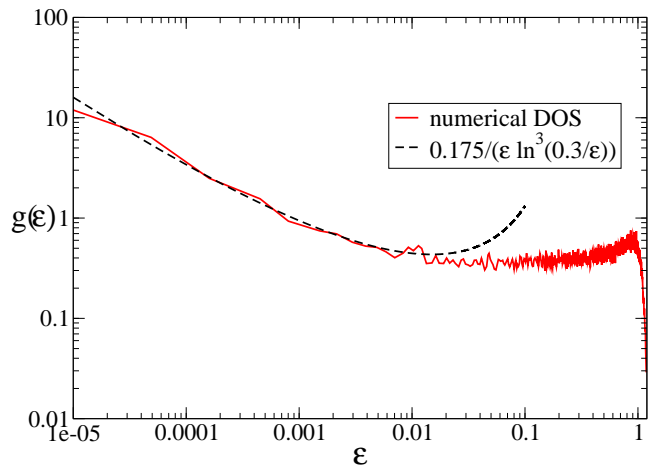}
\caption{Density of states of the random hopping problem with zero tilt 
for 2000 lattice sites, averaged over 50 disorder realizations. 
The hopping parameters $w_i$ are equally distributed in $[0.2, 0.7]$. 
At low energies, good agreement is obtained with the analytic RSRG result 
Eq.~(\ref{3.8}). \label{fig_DOS}}
\end{figure}

\begin{figure}
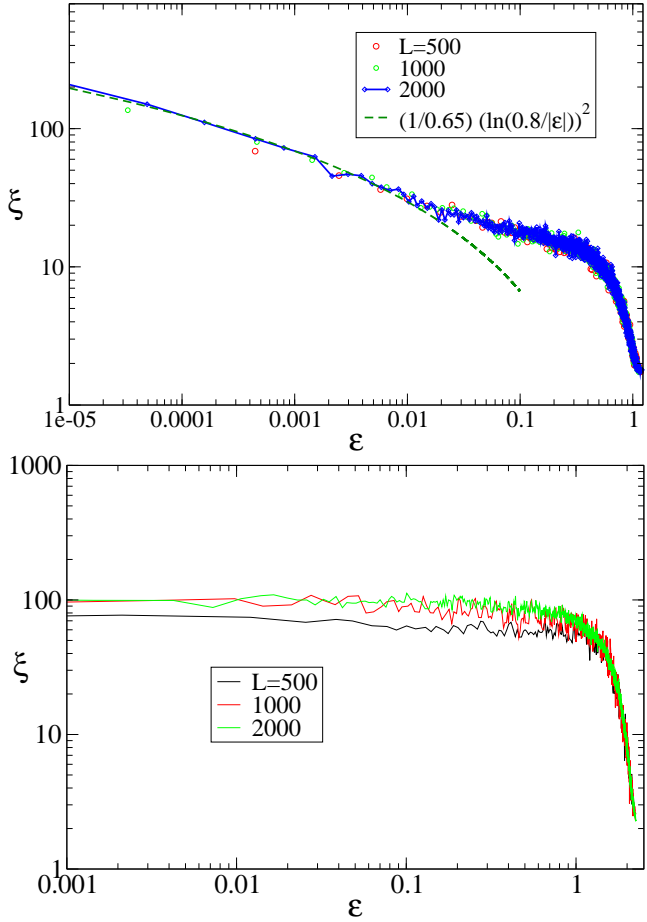

\includegraphics[width=8.5cm]{loc_length_bond_disorder_2_final} \hfill
\includegraphics[width=8.5cm]{loc_length_onsite_disorder_1_final}
\caption{Top: localization length of the random \emph{hopping} problem with zero tilt 
for different system sizes $L$, averaged over 50 disorder realizations. 
As in Fig.~\ref{fig_DOS}, the hopping matrix elements $w_i$ are equally distributed in $[0.2, 0.7]$.
The low-energy behavior agrees very well with the result  
$\xi \sim {1 \over \den_0} \ln^2(\Omega_I / |\epsilon|)$ predicted by the RSRG. 
Bottom: localization length for \emph{onsite} disorder with site energies $\epsilon_i$ 
equally distributed in $[-0.5, 0.5]$, averaged over 10 realizations. 
$\xi(\epsilon)$ clearly stays finite as $\epsilon \to 0$. 
\label{fig_loc_length}}
\end{figure}

\begin{figure}
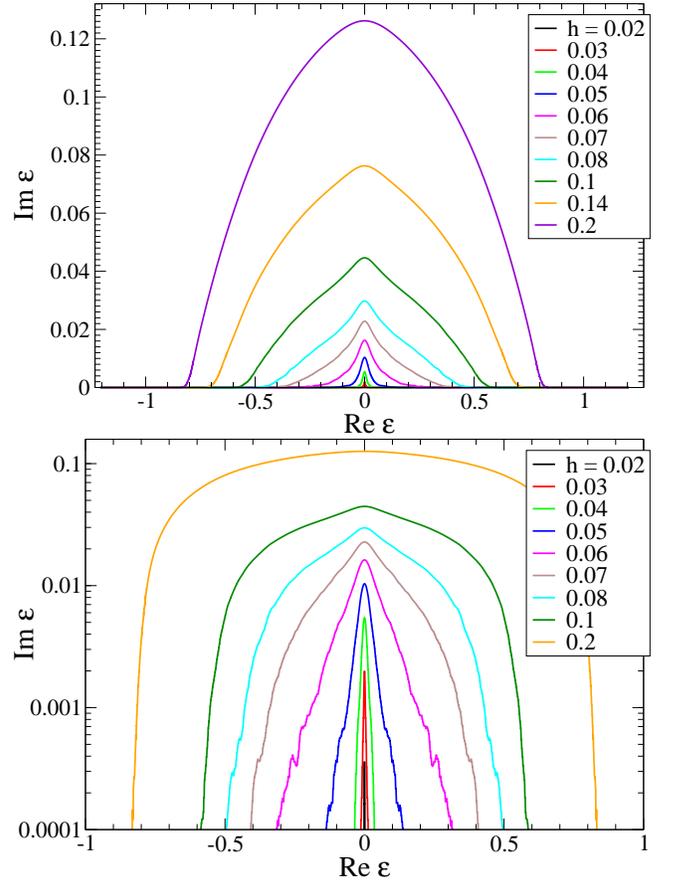

\includegraphics[width=8.5cm]{spectrum_random_hopping_II_linlin_final}
\hfill
\includegraphics[width=8.5cm]{spectrum_random_hopping_II_linlog_final}
\caption{Spectrum of single-particle states for different values of the tilt, with  
hopping disorder as in Fig.~\ref{fig_DOS}, $L=2000$ lattice sites and  
averaging over 100 disorder realizations, on a linear scale (top) and
in a lin-log plot (bottom). 
\label{fig_spectrum}}
\end{figure}

\begin{figure}
\includegraphics[width=8.5cm]{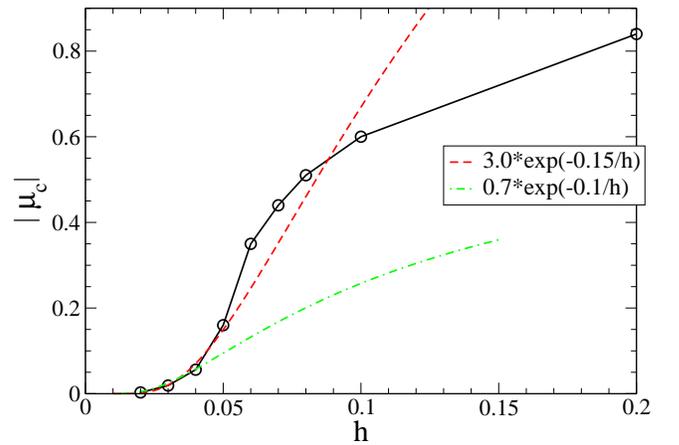}
\caption{Mobility edge vs. tilt for different values of the tilt, with  
hopping disorder as in Fig.~\ref{fig_DOS}, $L=2000$ lattice sites and  
averaging over 100 disorder realizations. 
\label{fig_mobility_edge}}
\end{figure}

\begin{figure*}
\includegraphics[width=13cm]{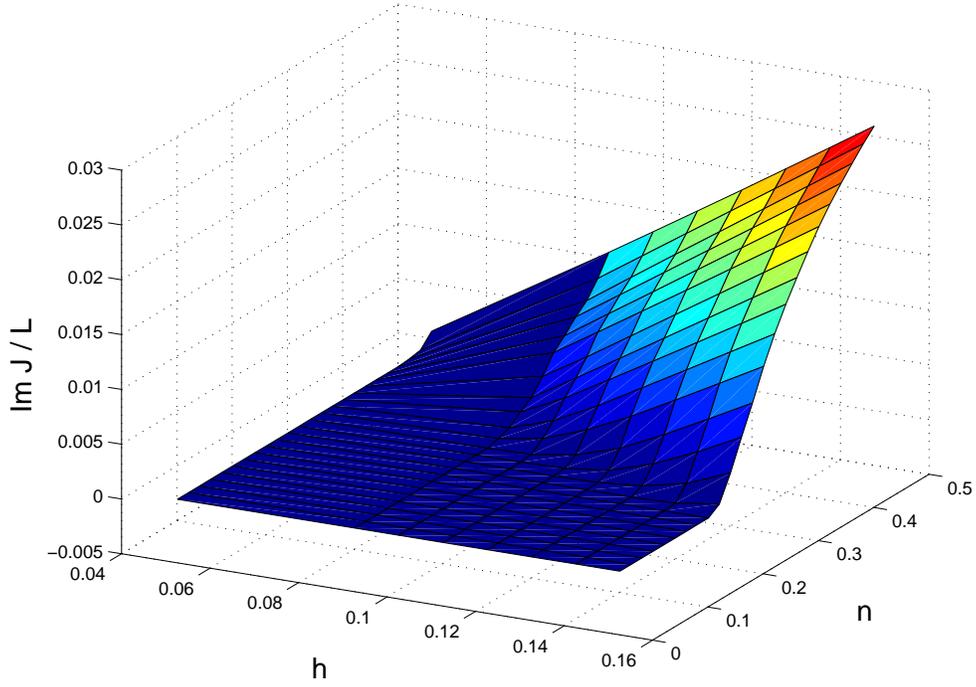}
\caption{Imaginary current for random hopping distributed as in Fig.~\ref{fig_DOS}, 
as a function of tilt $h$ and vortex density $n$. Results are given for $L=1000$ sites, 
with a disorder average performed over 20 realizations. 
\label{fig_imag_current}}
\end{figure*}

\begin{figure}
\includegraphics[width=8.5cm]{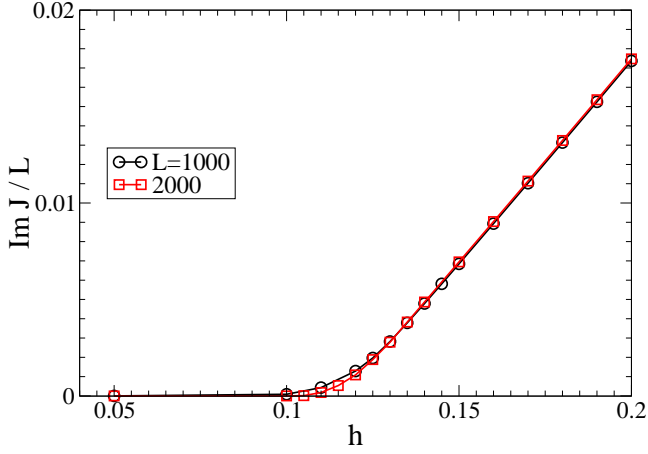}
\caption{Onset of imaginary current (i.e., transverse magnetic flux) as a function of tilt,  
for random hopping as in Fig.~\ref{fig_DOS} and vortex density $n=0.25$. 
Data are averaged over 20 realizations of disorder. 
Apart from finite-size rounding 
the onset is linear, as predicted by Eq.~(\ref{7.12}). 
\label{fig_imag_current_onset}}
\end{figure}

\section{Exact diagonalization of finite systems\lb{numerics}}

In this section we analyze the free-fermion Hamiltonian (\ref{1.7}) by exact numerical diagonalization of 
finite-size systems with up to $N=4000$ lattice sites. While this study is mostly focused on the case 
of random hopping, we also present some results for on-site disorder. 
We employ periodic boundary conditions in order to model qualitatively a realistic superconducting slab where vortices 
can enter and exit at the edges. Note that for open boundary
conditions (i.e., zero hopping matrix elements at the edges), the magnetic field tilt could be 
gauged away and would not induce any transverse vortex density. 
Since all quantities are random, we average over a finite number of disorder realizations, $N_{R} = 20 \ldots 100$.  
Unless indicated otherwise, the random hopping matrix elements $w_i$ are drawn from a flat 
distribution in $[c, c+0.5]$, where the lower cutoff $c=0.2$ is used to prevent singular couplings close to zero, 
which would lead to almost decoupled subsystems. For diagonal disorder, we use on-site energies $\epsilon_i$ 
which are evenly distributed in $[-0.5, 0.5]$. 

We first study the case of vanishing tilt $h=0$, in order to illustrate the qualitatively different 
physics of bond and on-site disorder. 
For bond disorder, we have calculated the density of states from the real part of the 
single-particle spectrum, averaging over $N_R = 50$ realizations of the disorder. 
The results are shown in Fig.~\ref{fig_DOS}, where it is evident that the DOS diverges at the center of the band, 
as predicted by the RSRG. We obtain a good fit to the theoretical result in Eq.~(\ref{3.8}).   

A similar behavior is obtained for the localization length $\xi$, which  we define as 
\be
\xi = \sum_{i,j=1}^L |\Psi_i |^2 |\Psi_j |^2 d_c(i,j)
\ee
where 
\be
d_c(i,j) = {L \over \pi} \sin\left( {\pi |i-j| \over L}\right) 
\ee
is the chord distance on the ring and $\Psi_i$ is the wave function at site $i$ of the single-particle state 
under consideration. 
Results are shown in Fig.~\ref{fig_loc_length}. 
For random hopping we find that the localization length diverges in the middle of the band 
according to 
\be
\xi=l\sim \frac{1}{\den_0} \ln^2 \frac{\Omega_I}{|\epsilon|}
\ee
as predicted by the RSRG [Eq. (\ref{3.71})]. 
In particular, there is always a vortex state at zero energy which is delocalized across the whole system. 
On the other hand, for on-site disorder $\xi$ clearly stays finite, as shown in the right graph 
of Fig.~\ref{fig_loc_length}. As a result, for bond disorder an arbitrarily small but finite tilt $h$ leads 
to extended states at the center of the band, in contrast to the case of on-site disorder where this only happens 
above a critical tilt $h>h_c$. \cite{HatanoNelson1997} [Note that the
  localization length discussed here is the distance between the two
  pinning sites that share a vortex, as opposed to $\xi_{\perp}$ of
  Sec. \ref{perp}, which describes the wondering of the vortex away
  from the pinning sites due to the tilt.]

We now present results for finite tilt $h>0$. In this case the Hamiltonian is non-Hermitian, 
with complex eigenvalues and left/right eigenstates which are no longer equal. 
In Fig.~\ref{fig_spectrum} we show numerical results for the spectrum of the random hopping problem. 
For small tilt almost all states are localized and the corresponding eigenenergies real. 
As the tilt increases, the two mobility edges $\pm \mu_c$ move towards the band edges and 
define a growing region of extended states. In Fig.~\ref{fig_mobility_edge}, the dependence of $\mu_c$ 
on the tilt is shown. The rapid vanishing of $\mu_c$ at small $h$ is
consistent with Eq. (\ref{3.83}) using Eq. (\ref{5.23}). 
(two fits are shown), 
although a quantitative comparison is difficult due to the finite system size. 

At finite tilt, the most interesting physical quantity for the vortex problem is the transverse flux, 
corresponding to the imaginary current $J$.   
Following Eq.~(\ref{current_op}), we obtain the current as the derivative of the ground-state energy $E_g$
with respect to tilt: 
\be
J = (-i) {1 \over L} {dE \over dh}
\ee
The resulting current for hopping disorder, as a function of particle (vortex) density $n$ and tilt $h$, 
is shown in Fig.~\ref{fig_imag_current}. 
Clearly, for weak tilt and few vortices, the current vanishes since the mobility edge is close to the 
center of the band and all the occupied states are localized. 
This is a manifestation of the transverse Meissner effect: In the presence of disorder due to 
columnar pins, a weak transverse magnetic field does not induce a transverse magnetic flux,  
because the flux lines are pinned by the defects.\cite{NelsonVinokur1993}
With increasing filling and tilt, a well-defined transition to 
finite current occurs when the mobility edge $\mu_c$ coincides with the vortex chemical 
potential $\mu$. This is shown in more detail in Fig.~\ref{fig_imag_current_onset}, where the current is 
given as a function of the tilt for quarter filling $n=0.25$. Apart from finite-size rounding of the transition, 
the onset of the current is linear in $h$, consistent with the theoretical prediction in Eq.~(\ref{7.12}).

\section{Summary}

In this paper we studied the physics of interacting vortices in a 2-dimensional type-II superconductor 
with parallel random columnar defects in the plane. 
Our analysis is based on mapping the system onto a 1+1-dimensional ensemble of hard-core bosons with 
on-site or hopping disorder. The equivalence of hardcore bosons and fermions in one spatial dimension 
allowed us to focus on an effectively noninteracting, but disordered,
many-fermion system for the special case of Luttinger liquid parameter
$g=1$. 
As a qualitative benchmark, we considered the exactly solvable Lloyd model with on-site disorder 
where we obtained analytic predictions for the critical tilt and the transverse magnetic flux above the 
critical tilt. We then studied extensively the case of general random
hopping, using the real-space renormalization group (RSRG) technique, which we generalized 
to the case of finite tilt. The RSRG allowed us to completely describe the localized phase and to 
extract physical quantities like the density of states or the mobility edge as a function of tilt. 
With the help of a non-Hermitian spectral formula we were able to extend the RSRG results 
to the delocalized states, and to calculate the onset of transverse magnetic flux close to the  
breakdown of the transverse Meissner effect, which is the most relevant experimental observable. 
Finally, we have compared our analytic predictions to numerics on finite systems which we 
diagonalized exactly. We found good agreement with the RSRG results, although the numerics indicates 
that the universal regime is only accessible for very large system sizes. 

Our results seem to indicate that scaling pictures formerly held
regarding the case of uniform pins at a random separation are not
valid near critical tilting, particularly, we find  $B_{\perp}\propto\l(H_{\perp}-H_{\perp}^c\r)^{\zeta}$
 in $1+1$ dimensions, but with $\zeta=1$ rather than the expected $\zeta=1/2$ (see
Secs. \ref{intro} and \ref{vcurrent}). We hope that our results will
encourage further experimental and theoretical research on vortex
pinning, both to verify our predictions, and to explore how strong
randomness physics may appear in situations not touched upon here.


\acknowledgements D.R.N. would like to thank I. Affleck for
discussions on torque measurements. D.R.N. was supported by the
National Science Foundation through grant No. DMR-0231631 and through
the Harvard Materials Research Science and Engineering Laboratory via
Grant No. DMR-0213805. G.R. Would like to thank the generous
hospitality of the Kavli Institute of Theoretical Physics, UCSB, and
of the Boston University visitors program.

\appendix

\section{A general formula for the transverse vortex density near the mobility edge\lb{appcurrent}}

If we increase the magnetic field above the critical field, some
vortices are delocalized, and a transverse vortex density appears. The
transverse vortex density (imaginary current) for a single vortex state is given by: 
\be
J_n=\frac{1}{L}{\rm Re}\der{E_n}{h}
\lb{7.5}
\ee  
Following Ref. \onlinecite{ShnerbNelson1998,Dahmen}, we write the eigenvalues of the
random hopping problem as
\be
\epsilon_n(h)=E(h+ik_n).
\lb{7.6}
\ee
The function $E(h+ik)$ is analytic and therefore obeys the
Cauchy-Riemann equations. In particular: 
\be
{\rm Re}\der{E}{h}={\rm Im}\der{E}{k}
\lb{7.7}
\ee
From Eq. (\ref{7.7}) we can get a general result for the total current:
\begin{widetext}
\be
J_{total}=2\frac{1}{L}\summ_{k_n}{\rm Re}\der{E_n}{h}\rightarrow
2\intt_{k_{min}}^{k_{max}}\frac{dk}{2\pi} {\rm Im}\der{\epsilon}{k}=\frac{1}{\pi}\l({\rm Im} E_{k_{max}}-{\rm Im} E_{k_{min}}\r)=\frac{1}{\pi}\l({\rm Im} E_{\mu}-{\rm Im} E_{\mu_m}\r)
\lb{7.8}
\ee
\end{widetext}
From Eq. (\ref{7.8}) we can immediately conclude, for instance, that
closed bubbles of tilted states don't contribute to the total current. The factor of two accounts for the two branches of the spectrum.

\section{The RSRG and critical tilt from the discrete non-Hermitian spectral formula \label{appA}}
 
The discrete non-Hermitian formula allows an alternative formulation
of the real-space RG, and thus gives further support to the results
obtained in Sec. \ref{results}. We will use the the real part of the
non-Hermitian spectral formula, Eq. (\ref{6.2}):
\be
\summ_{i=1}^{N}\frac{1}{2}\ln\l(({\rm Re} E-\epsilon_i)^2+({\rm Im} E)^2\r)=N|h|+ \summ_{i=1}^{N}
\ln\l(w_i\r).
\label{a1}
\ee
Because of particle-hole symmetry, we can also write:
\begin{widetext}
\be
\summ_{i=1}^{N/2}\frac{1}{2}\l(\ln\l(({\rm Re} E-\epsilon_i)^2+({\rm Im} E)^2\r)+\ln\l(({\rm Re} E+\epsilon_i)^2+({\rm Im} E)^2\r)\r)=N|h|+ \summ_{i=1}^{N}
\ln\l(w_i\r).
\label{a2}
\ee
\end{widetext}

The real-space RG is obtained as follows. Assuming strong disorder, as
we did in Sec. \ref{RSRGsec}, we know that the eigenvalues $\pm
\epsilon_1$ are associated with the strongest bond of the chain,
$n_1$:
\be
\epsilon_1\approx w_{n_1}
\lb{a3}
\ee
Let us choose an energy $|E|\ll \epsilon_1$, and assume that it
belongs to a localized state. We can now write:
\be
\summ_{i=2}^{N/2}\ln\l|E^2-\epsilon_i^2\r|=N|h|+
\summ_{i=1}^{N}
\ln\l(w_i\r)-\ln \epsilon_i^2.
\label{a4}
\ee
However, we can rearrange terms on the right hand side as follows:
\be
\ba{c}
\ln w_{n_1-1}+\ln w_{n_1}+\ln w_{n_1+1}-\ln \epsilon_i^2\vspace{2mm}\\
\approx
\ln \l(\frac{w_{n_1-1}w_{n_1+1}}{w_{n_1}}\r)=\ln{w^{eff}_{n_1-1,n_1+2}},
\lb{a5}
\ea
\ee
where $w^{eff}_{n_1-1,n_1+2}$ is just the effective hopping according
to the real space RG as found in Eq. (\ref{2.9}). We now have the
renormalized chain with the two sites $n_1,\,n_1+1$ removed, and we can
write the spectral formula as:
\be
\summ_{i=2}^{N/2}\ln\l|E^2-\epsilon_i^2\r|=N|h|+
\summ_{i=1}^{N-2}
\ln\l(\tilde{w}_i\r)-\ln \epsilon_i^2.
\label{a6}
\ee
where the $\tilde{w}_i$ are the hoppings of the renormalized
chain. This process can now be repeated - starting with the next eigenvalue, $\epsilon_2$ of the
largest renormalized bond, $\max\{\tilde{w}_i\}=\tilde{w}_{n_2}$,
the bond $n_2$ can be renormalized in the same way as written
above. Thus we recover the real-space RG - we can eliminate the high
energy states and write the spectral formula in terms of the new
couplings, and the reduced length. Note that the tilt part remains
constant in the process. This invariance corresponds  to the second condition in
Eq. (\ref{2.9}), which can be written as:
\be
\summ_i h_i=\summ_{\tilde{i}}\tilde{h}_{\tilde{i}}.
\lb{a7}
\ee
We can repeat this process until we reach energy eigenvalues of the
same order as $E$. 

We can also 
re-derive the critical tilt result, Eq. (\ref{4.5}). 
We assume that when close to the
critical tilt, or to the mobility edge, the eigenvalue $E$, which we assume is
$\epsilon_m<|E|<\epsilon_{m+1}$, is as far from
$\epsilon_m,\,\epsilon_{m+1}$ as it can be [such that the LHS of
Eq. (\ref{a6}) is maximized], and therefore:
\be
\frac{1}{N}\summ_{i\sim m}\ln||E|-\epsilon_i|\ll 1.
\lb{a8}
\ee
where $i\sim m$ indicates summing over $i$'s in the vicinity of $m$
and $m+1$. Eq. (\ref{a8}) is justified by the above assumption, {\it and} our understanding
that in the random singlet phase, when renormalizing a bond with
energy $E$, there is negligible probability to find another bond of
similar strength next to it. With Eq. (\ref{a8}), we can now write
  the spectral formula assuming we can carry out the RG procedure
  above until we arrive at the energy scale $|E|$. Also, since we
  ignore eigenvalues of the order of $|E|$, all remaining eigenvalues
  are much smaller than $|E|$. Therefore we can write: 
\be
\summ_{i=1}^{\tilde{N}/2}\ln\l|E^2\r|\approx N|h|+\summ_{i=1}^{\tilde{N}}\ln\l(\tilde{w}_i\r),
\label{a9}
\ee
where $\tilde{N}$ is the length of the chain renormalized to energy
scale $|E|$. This, in turn, can be written as:
\be
\summ_{i=1}^{\tilde{N}}\ln\frac{|E|}{\tilde{w}_i}=\tilde{N}\l(\G_E+\G_0\r)
\approx N h_c
\label{a10}
\ee
where we used Eq. (\ref{3.6}) to express the average of the
logarithmic couplings in terms of $\G_E$. We have now made the replacement 
$|h|\rightarrow h_c$, the critical tilt. Before, we assumed that the
tilt is as close to critical as possible, and therefore the LHS of
Eq. (\ref{a10}) is the maximum possible value for $N|h|$, without
letting $E$ have an imaginary part. To get $|h|<h_c$ we only need to
allow $E$ to come arbitrarily close to $\epsilon_m$. Thus we conclude:
\be
h_c=\frac{\tilde{N}\l(\G_E+\G_0\r)}{N}=\frac{\den_0}{\G_{\mu}+\G_0},
\lb{a11}
\ee
where we used Eq. (\ref{3.72}) for the effective density of the
renormalized chain. This concludes an independent demonstration of
Eq. (\ref{4.5}).

\bibliography{NHbib}

\end{document}